\documentclass[universe,review,accept,moreauthors,pdftex]{Definitions/mdpi} 

\firstpage{1} 
\pubvolume{1}
\issuenum{1}
\articlenumber{0}
\pubyear{2023}
\copyrightyear{2023}
\datereceived{13 January 2023} 
\daterevised{21 January 2023} 
\dateaccepted{6 February 2023} 
\datepublished{} 
\hreflink{https://doi.org/} 

\def\apss{{Astroph.\@ \& \@Space \@Science\ }}

\Title{Hubble Tension: The Evidence of New Physics}

\TitleCitation{Hubble Tension: The Evidence of New Physics}

\Author{Jian-Ping Hu
 $^{1}$\orcidA{} and Fa-Yin Wang $^{1,2,}$*\orcidB{}}

\AuthorNames{Jian-Ping Hu and Fa-Yin Wang}

\AuthorCitation{Hu, J.-P.; Wang, F.-Y.}

\address{%
$^{1}$ \quad School of Astronomy and Space Science, Nanjing University, Nanjing 210093, China; \mbox{hjp2022@nju.edu.cn}\\
$^{2}$ \quad Key Laboratory of Modern Astronomy and Astrophysics (Nanjing University), Ministry of Education, Nanjing 210093, China}

\corres{Correspondence: fayinwang@nju.edu.cn}

\abstract{The $\Lambda$CDM model provides a good fit to most astronomical observations but harbors large areas of phenomenology and ignorance. With the improvements in the precision and number of observations, discrepancies between key cosmological parameters of this model have emerged. Among them, the most notable tension is the 4$\sigma$ to 6$\sigma$ deviation between the Hubble constant ($H_{0}$) estimations measured by the local distance ladder and the cosmic microwave background (CMB) measurement. In this review, we revisit the $H_{0}$ tension based on the latest research and sort out evidence from solutions to this tension that might imply new physics beyond the $\Lambda$CDM model. The evidence leans more towards modifying the late-time universe.}


\begin{document}
\section{Introduction}\label{sec1}
The cosmological constant ($\Lambda$) cold dark matter model ($\Lambda$CDM) is the simplest cosmological model and consistent with the most astronomical observations \cite{2017MNRAS.470.2617A,2018ApJ...859..101S,2020A&A...641A...6P,2021A&A...647A..38B,2021MNRAS.507..730H,2021MNRAS.504.2535I,2022ApJ...938..110B,2022MNRAS.513.5517C,2022MNRAS.514.1828D,2022MNRAS.509.4745C,2022arXiv221104268D,2022ApJ...935....7L,2022PhRvD.106l3523P,2022ApJ...924...97W}. In the past \mbox{30 years}, it has been the best model, and no better one has yet been 
presented to replace it. Despite its remarkable successes, the validity of the $\Lambda$CDM model is currently under intense investigation \cite{2022PhRvD.106l3523P,2022arXiv220505017B,2022arXiv220306240B,2022arXiv220304757S} (for reviews, see \cite{2016IJMPD..2530007B,2022JHEAp..34...49A,2022Univ....8..399D,2022NewAR..9501659P,2022arXiv220705765A,2022arXiv220914918K}). The fine tuning and coincidence problems are the most important theoretical difficulties \cite{1989RvMP...61....1W,1997cpp..conf.....F,2012CRPhy..13..566M,2013arXiv1309.4133B,2014EPJC...74.3160V}. The fundamental problem with the former is that there is a large discrepancy between the observations and theoretical expectations of $\Lambda$ \cite{1989RvMP...61....1W,2006IJMPD..15.1753C,2012CRPhy..13..566M,2013JPhCS.453a2015S}. The latter is related to the observed vacuum energy density $\Omega_{\Lambda}$ and the matter energy density $\Omega_{m}$, which are now nearly equal despite their dramatically different evolutionary properties. The anthropic principle, as a possible solution to these problems, states that these ``coincidences'' results from a selection bias towards the existence of human life in the context of a multiverse \cite{1987PhRvL..59.2607W,2003dmci.confE..26S}. In addition to the above theoretical challenges, the two main components in the $\Lambda$CDM model, dark matter (DM) and dark energy (DE), are poorly understood. Moreover, there are also some tensions between the cosmological and astrophysical observations and the $\Lambda$CDM model, which include the Hubble tension  \cite{2016A&A...594A..13P,2018ApJ...855..136R,2019ApJ...876...85R,2020A&A...641A...6P,2020NatRP...2...10R,2020MNRAS.498.1420W,2021MNRAS.502.2065D,2021ApJ...908L...6R,2022ApJ...934L...7R} (5$\sigma$), growth tension \cite{2017PhRvD..96f3517B,2018PhRvD..98d3526A,2018MNRAS.474.4894J} (2-3$\sigma$), CMB anisotropy \mbox{anomalies \cite{2020PhRvD.102l4059A,2020A&A...643A..93H,2020A&A...641A...7P,2021A&A...649A.151M,2021ApJ...908L..51S,2021EPJC...81..694Z,2022JPhCS2191a2001A,2022arXiv221011333K,2022MNRAS.511.5661Z,2022arXiv220705765A,2023PDU....3901162A}} (2-3$\sigma$), cosmic dipoles \cite{2008ApJ...686L..49K,2009MNRAS.392..743W,2011PhRvL.107s1101W,2012MNRAS.422.3370K,2013PhRvD..88h3529W,2018JCAP...04..031B,2021EPJC...81..948Z,2022A&A...668A..34H,2022PhRvD.105j3510L,2022arXiv221204925G,2022arXiv220914918K} (2-5$\sigma$), Baryon Acoustic Oscillation (BAO) curiosities \cite{2017JCAP...04..024E,2018ApJ...853..119A,2019JCAP...10..044C} (2.5-3$\sigma$), parity violating rotation of CMB linear \mbox{polarization \cite{2019arXiv190412440M,2020PTEP.2020f3E01M,2020PhRvL.125v1301M,2020PTEP.2020j3E02M}}, small-scale curiosities \cite{2017ARA&A..55..343B,2017Galax...5...17D,2019A&ARv..27....2S,2020arXiv200503520D}, age of the universe \cite{2013PDU.....2..166V}, the Lithium problem \cite{2011ARNPS..61...47F} (2-4$\sigma$),the quasar Hubble diagram \cite{2019A&A...628L...4L,2019NatAs...3..272R,2020PhRvD.102l3532Y,2021PhLB..81836366B,2022A&A...661A..71H} ($\sim$4$\sigma$), oscillating signals in short range gravity experiments \cite{2017PhRvD..96j4002A,2017PhRvD..95h4050P}, anomalously low baryon temperature \cite{2018Natur.555...67B} ($\sim$3.8$\sigma$), colliding clusters with high velocity \cite{2015JCAP...04..050K,2021MNRAS.500.5249A} ($\sim$6$\sigma$), etc. More detailed information of these tensions can be found from \citet{2022NewAR..9501659P}. The $H_{0}$ tension emerged with the first release of \emph{Planck} results \cite{2014A&A...571A..16P}, and has grown in significance in the past few years \cite{2016A&A...594A..13P,2018ApJ...855..136R,2019ApJ...876...85R,2020A&A...641A...6P,2020NatRP...2...10R,2020MNRAS.498.1420W,2021MNRAS.502.2065D,2021ApJ...908L...6R,2022ApJ...934L...7R}. 

There have been many studies dedicated to finding out what causes the Hubble tension, but so far there is 
 no convincing explanation. The reanalyses of the Planck observations and Hubble Space Telescope (HST) measurements demonstrate that the serious discrepancy of $H_{0}$ may not be caused by systematics (including photometric biases, environmental effects, calibration error, lens mass modelling biases, CMB foreground effects and so \mbox{on) \cite{2017A&A...607A..95P,2018ApJ...867..108J,2019ApJ...876...85R,2019MNRAS.484L..64S,2020A&A...641A...1P,2020A&A...641A...7P,2020A&A...644A.176R,2022EPJP..137..537C,2022MNRAS.514.4620D,2022ApJ...934L...7R}}. Hence, many researchers prefer to believe that the $H_{0}$ tension might be caused by new physics beyond the $\Lambda$CDM model \cite{2020NatRP...2...10R}. So far, there has been a large number of modified models which adopted to resolve or relieve the $H_{0}$ tension, \mbox{see \cite{2022JHEAp..34...49A,2022NewAR..9501659P}} for a review of $H_{0}$ tension solutions. Although many extensions of $\Lambda$CDM can alleviate the $H_{0}$ tension, none are supported by the majority of 
 observations \cite{2020PhRvD.102d3507H,2021JCAP...05..072D}. There are many international conferences held on the $H_{0}$ tension, such as ``Beyond $\Lambda$CDM'', ``Hubble Tension Headache'', ``Tensions between the Early and the Late Universe'', etc. \cite{2021A&ARv..29....9S}. ``Beyond $\Lambda$CDM'' in Oslo in 2015, a poll at the conference showed that 69$\%$ of participants believe that new physics is the most likely explanation. On the contrary, more than 50$\%$ of the participants of the ``Hubble Tension Headache'' conference held by the University of Southampton supported the explanation that there were still systematics unknown to us in the observational data. Theoretical Physics workshop (``Tensions between the Early and the Late Universe'') in July 2019 directed attention to the Hubble constant discrepancy. The workshop pointed out that ``New results showed that it does not appear to depend on the use of any one method, team or source'' \cite{2019NatAs...3..891V}. To streamline the interaction between these different communities and promote the transparent transfer of information, participants also gave some reasonable suggestions. More details about the workshop are available at website: https://www.kitp.ucsb.edu/activities/enervac-c19. In a word, the research and discussion on the $H_{0}$ tension is still going on. 

This review is organized as follows. In Section \ref{sec2}, we briefly introduce the $H_{0}$ tension, and then detail two methods for constraining $H_{0}$, the CMB measurements and the local distance ladder. In Section \ref{sec3}, we discuss recent methods for estimating $H_{0}$ using otherwise independent observations. We also discuss a taxonomy of solutions to the $H_{0}$ tension in Section \ref{sec4}. Afterwards, from all the solutions to the $H_{0}$ tension, we sort out evidence that might imply new physics beyond the $\Lambda$CDM model and briefly introduce some of the proposed scenarios in Section \ref{sec5}. Finally, we give a brief conclusion and future prospects. 

\section{\boldmath{$H_0$} Tension} \label{sec2}
The local expansion rate of the universe $H_{0}$ is a fundamental value. It also determines the age of the universe; thus, it is important to determine it accurately. The accuracy of $H_{0}$ measurements has been improved as the number of probes has increased. A review of most well-established probes can be found in literature \cite{2013PhR...530...87W}. In general, the $H_{0}$ measurements can be estimated from the cosmological model utilizing the early universe measurements, or more directly measured from the local universe. Interestingly, the $H_{0}$ values measured by these two approaches are inconsistent. The discrepancy between the $H_0$ values measured from the local distance ladder and from the CMB is the most serious challenge to the standard $\Lambda$CDM model. This $H_{0}$ discrepancy is also known as ``Hubble Tension'' \cite{2014A&A...571A..16P}. 

\subsection{{Constraints of $H_{0}$ from the CMB Measurements}}
The estimation of $H_{0}$ from the CMB data proceeds in three steps \cite{2013ApJS..208...20B,2020PhRvD.101d3533K}: (1) determine the baryon density and matter density to allow for the calculation of the sound horizon size ($r_{s}$); (2) infer $\theta_{s}$ from the spacing between the acoustic peaks to determine the comoving angular diameter distance to last scattering $D_{A}$ = $r_{s}$/$\theta_{s}$; (3) adjust the only remaining free density parameter in the model that $D_{A}$ gives this inferred distance. With this last step complete, we now have $H(z)$ determined for all $z$, including $H_{0}$ (z = 0). 

In 2013, \citet{2013ApJS..208...20B} provided $H_{0}$ = 70.00 $\pm$ 2.20 km/s/Mpc through analysing the nine-year  Wilkinson Microwave Anisotropy Probe (WMAP) observations. Meanwhile, the first data release of the \emph{Planck} space observatory, which was operated by the European Space Agency (ESA), gave a precise result $H_{0}$ = 67.40 $\pm$ 1.40 km/s/Mpc \cite{2014A&A...571A..16P}. After that, a more accurate $H_{0}$ = 67.40 $\pm$ 0.50 km/s/Mpc yielded by the \emph{Planck} final data release is also in line with the \emph{Planck}2013 results \cite{2020A&A...641A...6P}. Researchers also consider adding the other observational data to constrain $H_{0}$, \emph{Planck}2018+lensing 67.36 $\pm$ 0.54 km/s/Mpc and \emph{Planck}2018+lensing+BAO 67.66 $\pm$ 0.42 km/s/Mpc \cite{2020A&A...641A...6P}. The main $H_{0}$ measurements are shown in Figure~\ref{fig1}. In addition, there exists a lot of $H_{0}$ predictions adopting other CMB experiments from the ground, including the South Pole Telescope (SPTPol) \cite{2019JCAP...10..044C,2021PhRvD.104b2003D} and Atacama Cosmology Telescope (ACT) \cite{2020JCAP...12..047A}. These $H_{0}$ predictions are all consistent with the \emph{Planck}2018 result \cite{2020A&A...641A...6P}. 

\subsection{Constrain $H_{0}$ from the Local Distance Ladder }
At present, there has been a lot of methods in the local universe to estimate the Hubble constant based on the distance-redshift relation. These methods are usually undertaken by building a ``local distance ladder''; the most common approach is to adopt geometry to calibrate the luminosity of specific star types. Cepheid variables are often employed to determine the distance of 10$-$40 Mpc \cite{2001ApJ...553...47F,2012ApJ...758...24F}. Measuring longer distances requires other standard candles, for example SNe Ia, 
 whose maximum redshift can reach 2.36 \cite{2018ApJ...859..101S,2022ApJ...938..110B}. Quasars \cite{2016ApJ...819..154L,2017FrASS...4...68B,2017A&A...602A..79L,2019MNRAS.489..517M,2020MNRAS.492.4456K,2022MNRAS.516.1721C,2022MNRAS.513.1985K,2022MNRAS.515.3729K,2022MNRAS.510.2753K,2022ApJ...940..174W} and gamma radio bursts (GRBs) \cite{2008MNRAS.391L..79D,2015NewAR..67....1W,2017NewAR..77...23D,2021MNRAS.507..730H,2022MNRAS.510.2928C,2022MNRAS.512..439C,2022MNRAS.509.4745C,2022MNRAS.513.5686C,2022arXiv221201990D,2022MNRAS.516.2575J,2022ApJ...931...50L,2022ApJ...941...84L,2022arXiv221214291L,2022arXiv220813700M,2022ApJ...924...97W,2023MNRAS.518.2247L} offer the prospect of extending the Hubble diagram up to higher redshifts.

The first $H_{0}$ estimation, using the Cepheid variables and SNe Ia provided by the HST project, was $72\pm8$ km/s/Mpc \cite{2001ApJ...553...47F}. A improved result, 74.3 $\pm$ 2.1 km/s/Mpc, was yielded by using a modified distance calibration \cite{2012ApJ...758...24F}. After that, the SH0ES project, which started in 2005, also produces many $H_{0}$ results \cite{2016ApJ...826...56R,2018ApJ...855..136R,2019ApJ...876...85R,2021ApJ...908L...6R}. Based on the SH0ES data, $H_{0}$ results from numerous reanalyses using different formalisms, statistical methods of inference, or replacement of parts of the data-set are both in line with the previous results \cite{2017JCAP...03..056C,2018ApJ...869...56B,2018MNRAS.476.3861F,2018MNRAS.477.4534F,2018A&A...609A..72D,2020PhRvR...2a3028C,2021ApJ...911...12J}. The latest $H_{0}$ result provided by the SH0ES collaboration shows that $H_{0}$ = 73.04 $\pm$ 1.04 km/s/Mpc \cite{2022ApJ...934L...7R}. 

\begin{figure}[H]
\includegraphics[width=6 cm]{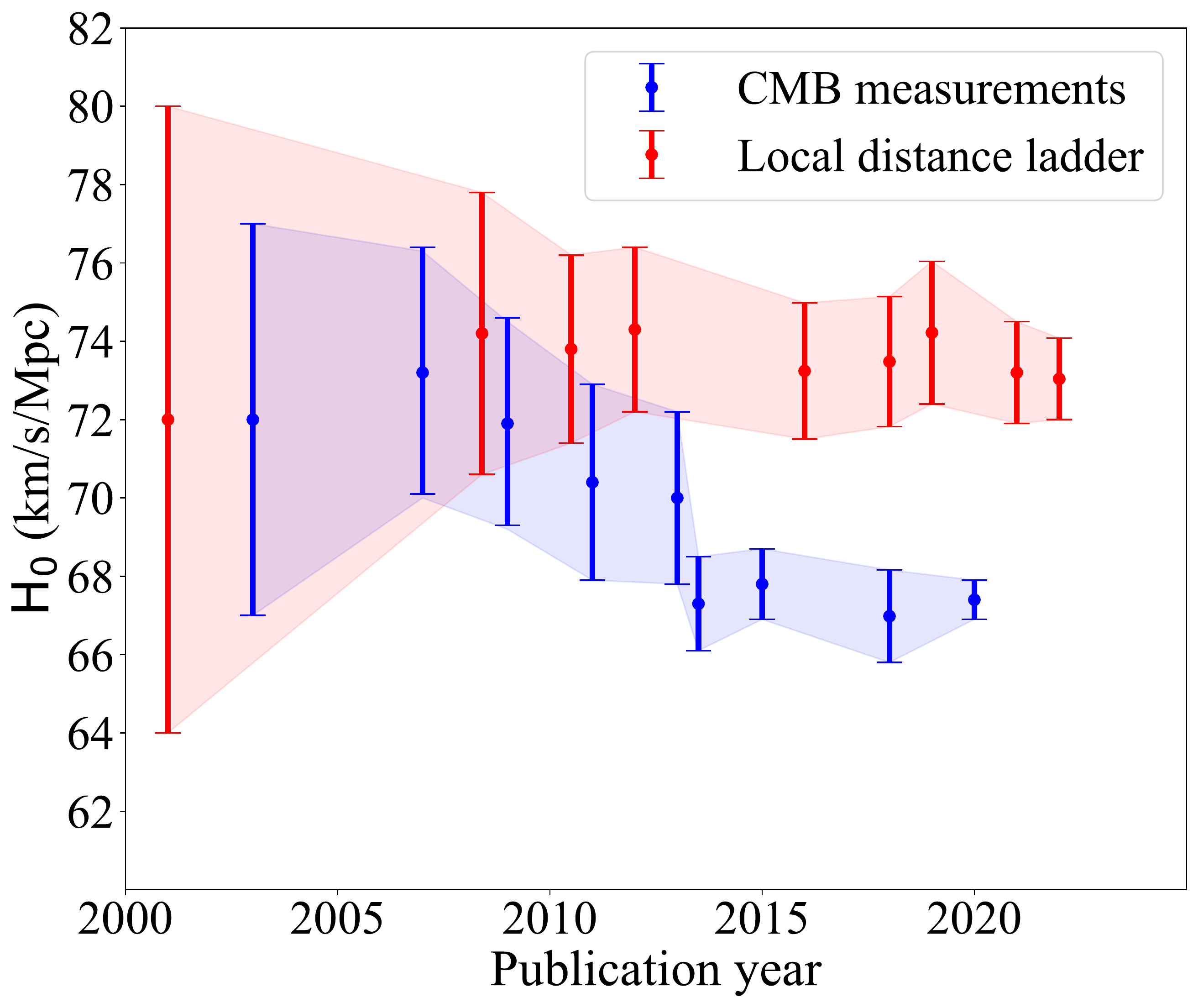}
\caption{$H_{0}$ measurements from the early-time and the late-time observations. Blue points denote $H_{0}$ estimations from analyses of CMB data, including first year WMAP \cite{2003ApJS..148..175S}, three year WMAP \cite{2007ApJS..170..377S}, five year WMAP \cite{2009ApJS..180..330K}, seven-year WMAP \cite{2011ApJS..192...18K}, nine-year WMAP \cite{2013ApJS..208...20B}, Planck 2013 \cite{2014A&A...571A..16P}, Planck 2015 \cite{2016A&A...594A..13P}, Planck 2018 \cite{2020A&A...641A...6P} and BAO \cite{2018ApJ...853..119A}. Red points denote $H_{0}$ values measured by utilizing the local distance ladder including the Cepheid distance scale \cite{2001ApJ...553...47F}, Carnegie Hubble Program \cite{2012ApJ...758...24F} and SH0ES \cite{2009ApJ...699..539R,2011ApJ...730..119R,2016ApJ...826...56R,2018ApJ...855..136R,2019ApJ...876...85R,2021ApJ...908L...6R,2022ApJ...934L...7R}. (Source: Figure 12 in \citet{2022NewAR..9501659P})}
\label{fig1}
\end{figure}

During the last few decades there has been remarkable progress in measuring the Hubble constant. The available technology and measurement methods determine the accuracy of this quantity. The main $H_{0}$ results up to 2022 obtained from the CMB measurements and the local distance ladder, as a function of the publication year, 
 is shown in Figure~\ref{fig1}. The uncertainties in these values have been decreasing for both methods and the recent measurements disagree by 5$\sigma$ \cite{2022ApJ...934L...7R}. There is an obvious $H_{0}$ tension without \mbox{physical explanations}. 

\unskip

\section{\boldmath{$H_{0}$} Arbitration}\label{sec3}
The \emph{Planck} CMB measurements and the local ladder measurements both provide very precise constraints on cosmological parameters. However, as with any experimental measurement, they are not free from systematic errors. Hence, possible systematics in the Planck observations and the HST measurements are suspected to be responsible for the Hubble discrepancy in the early and late measurements. However, this possibility has been largely ruled out \cite{2019ApJ...876...85R,2017A&A...607A..95P,2018ApJ...867..108J,2019MNRAS.484L..64S,2020A&A...641A...1P,2020A&A...641A...7P,2020A&A...644A.176R,2022ApJ...934L...7R,2022MNRAS.514.4620D}. Since reanalyses of the \emph{Planck} and local ladder measurements could not find a satisfactory answer to the $H_{0}$ tension, hope was pinned on other observations, such as quasar lensing \citep{2020MNRAS.498.1420W,2020AA...639A.101M}, Megamaser  \cite{2013ApJ...767..155K,2013ApJ...767..154R,2015ApJ...800...26K,2019ApJ...886L..27R,2020ApJ...891L...1P}, gravitational waves (GW) \cite{2017Natur.551...85A,2018Natur.561..355M,2019NatAs...3..940H}, fast radio bursts (FRBs) \cite{2022MNRAS.511..662H,2022MNRAS.515L...1W,2022MNRAS.516.4862J,2022arXiv221005202L,2022arXiv221213433Z}, tip of the red giant branch (TRGs)~\cite{2019ApJ...882...34F,2020ApJ...891...57F,2021ApJ...919...16F}, BAOs \cite{2018ApJ...853..119A,2021A&A...647A..38B}, Type II supernovae \cite{2022MNRAS.514.4620D}, Ages of Old Astrophysical \mbox{Objects \cite{2022JHEAp..36...27V,2022ApJ...928..165W}}, etc. \cite{2022LRR....25....6M}. These observations do not assume a cosmological model and are independent of the CMB and distance ladder measurements. 

\subsection{Quasar Lensing}
Quasar lensing as an independent method can be used to estimate $H_{0}$. With observed time delay $\Delta \tau_{obs}$ and lens mass model, $H_{0}$ can be inferred. The observed time delay is owing to the geometrical path length difference caused by the gravitational potential of the lens, which is associated with the path of the light rays from the vicinity of the lens to the observer \cite{2002LNP...608.....C}. The time delay distance $D_{\Delta t}$ inferred from $\Delta \tau_{obs}$ is actually a combination of angular diameter distances: \cite{2020MNRAS.498.1420W}
\begin{eqnarray}\label{Dt}
	D_{\Delta t} \equiv (1+z_{lens})\frac{D_{d}D_{s}}{D_{ds}},
\end{eqnarray}
where $z_{lens}$ is the lens redshift, $D_{d}$ is the angular diameter distance to the lens, $D_{s}$ is the angular diameter distance to the source, and $D_{ds}$ is the angular diameter distance between the source and the lens. In the flat $\Lambda$CDM model, $D_{ds}$ can be given as \cite{2018SSRv..214...91S}
\begin{eqnarray}\label{Dds}
	D_{ds} = \frac{c}{H_{0}(1+z_s)}\int_{z_{lens}}^{z_s} (\Omega_{m}(1+z')^{3}+ \Omega_{\Lambda})^{-1/2} dz',
\end{eqnarray}
here, $z_s$ is the redshift of source. Substitute $z_{lens}$ and $z_s$ into the following formula:
\begin{eqnarray}\label{eq:D1}
	D_{i} = \frac{c}{H_{0}(1+z_i)}\int_{0}^{z_i} \sqrt{\Omega_{m}(1+z')^{3}+ \Omega_{\Lambda}} dz',
\end{eqnarray}
the expressions of $D_{d}$ and $D_{s}$ can be obtained. The time delay distance is primarily sensitive to $H_{0}$, with weak dependence on other cosmological parameters. There were seven systems which provided the estimations of $H_{0}$, as shown in the following Table~\ref{tab1}. In addition, \citet{2020MNRAS.498.1420W} and \citet{2020AA...639A.101M} utilize the multiple lens systems to constrain $H_{0}$; the former provides a result of $73.3^{+1.7}_{-1.8}$ km/s/Mpc and the latter provides a result of $74.0^{+1.7}_{-1.8}$ km/s/Mpc. Recently, \citet{2023arXiv230102656S} updated the $H_{0}$ measurement ($78.3^{+3.4}_{-3.3}$ km/s/Mpc) derived from the lens RXJ1131-1231, and provided a new $H_{0}$ measurement ($74.2^{+1.6}_{-1.6}$ km/s/Mpc) for seven lenses. Their new measurement is in excellent agreement with those obtained in the past using standard simply parametrized mass profiles.

\begin{table}[H] 
\caption{Summary of the $H_{0}$ estimations derived from the quasar lensing systems.\label{tab1}}
\newcolumntype{C}{>{\centering\arraybackslash}X}
\begin{tabularx}{\textwidth}{p{2.7cm}p{2cm}p{2cm}p{2.5cm}p{2cm}}
\toprule
\textbf{Lens Name} & \boldmath{$z_{d}$} & \boldmath{$z_{S}$}		& \boldmath{$H_{0}$} \textbf{(km/s/Mpc)}	& \textbf{Reference}\\
\midrule
B1608+656   	 & 0.6304	& 1.394	 & $71.0^{+2.9}_{-3.3}$			& \cite{2010AA...524A..94S,2019Sci...365.1134J}\\
RXJ1131-1231	 & 0.295	& 0.654  & $78.3^{+3.4}_{-3.3}$			        & \cite{2014ApJ...788L..35S,2019MNRAS.490.1743C,2023arXiv230102656S}\\
HE0435-1223		 & 0.4546	& 1.693	 & $71.7^{+4.8}_{-4.5}$			& \cite{2017MNRAS.465.4895W,2019MNRAS.490.1743C}\\
SDSS 1206+4332	 & 0.745	& 1.789	 & $68.9^{+5.4}_{-5.1}$			& \cite{2019MNRAS.484.4726B}\\
WFI2033-4723	 & 0.6575	& 1.662	 & $71.6^{+3.8}_{-4.9}$   		& \cite{2020MNRAS.498.1440R}\\
PG1115+080		 & 0.311	& 1.722  & $81.1^{+8.0}_{-7.1}$		    & \cite{2019MNRAS.490.1743C}\\
DES J0408-5354	 & 0.597    & 2.375  & $74.2^{+2.7}_{-3.0}$		    & \cite{2017ApJ...838L..15L,2020MNRAS.494.6072S}\\
\bottomrule
\end{tabularx}
\end{table}
\unskip

\subsection{Megamaser}
Water megamasers residing in the accretion disks around supermassive black holes (SMBHs) in active galactic nuclei (AGNs) provide a unique way to bypass the distance ladder and make one-step, geometric distance measurements to their host galaxies \cite{2020ApJ...891L...1P}. The archetypal AGN accretion disk megamaser system is located in the nearby (7.6 Mpc) \cite{2013ApJ...775...13H,2019ApJ...886L..27R} Seyfert 2 galaxy NGC 4258 \cite{1984Natur.310..298C,1993Natur.361...45N,1999Natur.400..539H}. The Megamaser Cosmology Project (MCP) is a multi-year campaign to find, monitor, and map AGN accretion disk megamaser systems \cite{2007IAUS..242..399B,2008ApJ...678...96B}. The primary goal of the MCP is to constrain $H_{0}$ to a precision of several percent through making geometric distance measurements to megamaser galaxies in the Hubble flow \cite{2013ApJ...767..155K,2013ApJ...767..154R,2015ApJ...800...26K,2017ApJ...834...52G}. Distance measurements using the megamaser technique do not rely on the CMB or distance ladders measurements. Therefore, megamaser measurements provide an independent handle on $H_{0}$ estimates.

The MCP has so far published distances to six galaxies including UGC 3789 \cite{2013ApJ...767..154R}, NGC 6264 \cite{2013ApJ...767..155K}, NGC 6323 \cite{2015ApJ...800...26K}, NGC 5765b \cite{2016ApJ...817..128G}, NGC 4258 \cite{2019ApJ...886L..27R} and CGCG 074-064 \cite{2020ApJ...890..118P}. The latest details of these six galaxies are shown in Table~\ref{tab2}. In 2018, \citet{2018IAUS..336...86B} updated the published value to the UGC 3789 and obtained the $H_{0}$ estimation by using the first four galaxies, $H_{0}$ = 69.3$\pm$4.2 km/s/Mpc. Recently, \citet{2020ApJ...891L...1P} applied an improved approach for fitting maser data and obtained better distance estimates for the first four galaxies. Combining all the distance measurements of galaxies and assuming a fixed velocity uncertainty of 250 km/s in connection with peculiar motions, they provided the $H_0$ estimation, $H_{0}$ = 73.9$\pm$3.0 km/s/Mpc, independent of the CMB and distance ladders measurements.

\begin{table}[H] 
\caption{The details of six galaxies. The $H_{0}$ estimations are obtained by assuming a fixed velocity uncertainty of 250 km/s. (Source: Tables~1 and 2 in \citet{2020ApJ...891L...1P}.)\label{tab2}}
\newcolumntype{C}{>{\centering\arraybackslash}X}
\begin{tabularx}{\textwidth}{p{2.7cm}p{2.2cm}p{2.2cm}p{2.2cm}p{2cm}}
\toprule
\textbf{Galaxy Name} & \textbf{Distance (Mpc)} & \textbf{Velocity (km/s)}	& \boldmath{$H_{0}$} \textbf{(km/s/Mpc)}	& \textbf{Reference}\\
\midrule
UGC 3789   	 & $51.5^{+4.5}_{-4.0}$	& 3319.9 $\pm$ 0.8       & $75.8^{+3.4}_{-3.3}$	    & \cite{2013ApJ...767..154R}\\
NGC 6264	 & $132.1^{+21.0}_{-17.0}$	& 10192.6 $\pm$ 0.8  & 73.8$^{+3.2}_{-3.2}$	    & \cite{2013ApJ...767..155K}\\
NGC 6323	 & $109.4^{+34.0}_{-23.0}$	& 7801.5 $\pm$ 1.5   & $73.8^{+3.1}_{-3.0}$	    & \cite{2015ApJ...800...26K}\\
NGC 5765b    & $112.2^{+5.4}_{-5.1}$	& 8525.7 $\pm$ 0.7   & $74.1^{+4.5}_{-4.4}$		& \cite{2016ApJ...817..128G}\\
NGC 4258	 & $7.58\pm0.11$    	& 679.3 $\pm$ 0.4	       & $73.6^{+3.1}_{-3.0}$   	& \cite{2019ApJ...886L..27R}\\
CGCG 074-064 & $87.6^{+7.9}_{-7.2}$	& 7172.2 $\pm$ 1.9       & $72.5^{+3.4}_{-3.2}$	    & \cite{2020ApJ...890..118P}\\
\bottomrule
\end{tabularx}
\end{table}

\subsection{Gravitational Wave}
On 17 August 2017, the Advanced Laser Interferometer Gravitational-wave Observatory (LIGO) \cite{2015CQGra..32g4001L} and Virgo \cite{2015CQGra..32b4001A} detectors observed GW170817, a strong signal from the merger of a binary neutron-star system. Less than 2 seconds after the merger, a gamma-ray burst (GRB 170817A) was detected within a region of the sky consistent with the LIGO-Virgo-derived location of the GW source \cite{2017ApJ...848L..13A,2017ApJ...848L..14G,2017ApJ...848L..15S}. The detection of GW170817 in both gravitational waves and the electromagnetic (EM) waves heralds the age of the gravitational-wave multi-messenger astronomy. 

With the luminosity distance fitted from the GW waveform and the redshift information from the host galaxy, GW can be treated as standard sirens to conduct research in cosmology. The GW amplitude depends on the chirp mass and luminosity distance of the GW source. The mass can be precisely determined by the phase measurement of the GW signal. Therefore, as long as the amplitude and phase information of the GW source are obtained at the same time, the corresponding luminosity distance can be given. The heliocentric redshift measurement, $z_{helio}$ = 0.009783, was obtained from the optical identification of the host galaxy NGC 4993 \cite{2017ApJ...848L..12A}. The joint analysis of the GW signal from GW170817 and its EM localization led to the first $H_{0}$ estimation, $H_{0}$ = $74_{-8}^{+16}$ km/s/Mpc (median and symmetric 68$\%$ credible interval) \cite{2017Natur.551...85A}. In this analysis, the degeneracy in the GW signal between the source distance and the observing angle dominated the uncertainty of the $H_{0}$ measurement. Tight constraints 
 on the observing angle using high angular resolution imaging of the radio counterpart of GW170817 have been obtained \cite{2018Natur.561..355M}. \citet{2019NatAs...3..940H} reported an improved measurement of $H_{0}$ = $70.30_{-5.0}^{+5.3}$ km/s/Mpc by using these new radio observations, combined with the previous GW and EM data. Recently, using 47 GW sources from the Third LIGO-Virgo-KAGRA Gravitational-Wave Transient Catalog (GWTC-3), \citet{2021arXiv211103604T} presented $H_{0} = 68^{+12}_{-7}$ km/s/Mpc (68$\%$ credible interval) when combined with the $H_0$ measurement from GW170817 and its EM counterpart. Moreover, combining the GWTC-3 with the $H_0$ measurement from GW170817, \citet{2022arXiv220303643M} provided a more compact $H_{0}$ result of $67^{+6.3}_{-3.8}$ km/s/Mpc.

\subsection{Fast Radio Burst}
FRBs are millisecond-duration pulses occurring at cosmological distances \cite{Lorimer2007,Xiao2021,Zhang2022}. The total dispersion measure ($DM_{obs}$) of FRBs can provide a distance estimation to the source. The expanded form of $DM_{obs}$ is as follows:
\begin{eqnarray}\label{DM}
	DM_{obs}(z)  = DM_{MW} + DM_{IGM}(z) + \frac{DM_{host}(z)}{1+z},
\end{eqnarray}
where $DM_{IGM}$ represents the contribution of the intergalactic medium (IGM), $DM_{MW}$ are contributed by the interstellar medium (ISM) and the halo of the Milky Way, and the $DM_{host}$ is contributed by the host galaxy. Considering a flat universe, the averaged value of $\rm DM_{IGM}$ is \cite{deng14}
\begin{equation}\label{dmigm}
\langle {\rm{DM_{IGM}}(z)}\rangle = \frac{A \Omega_bH_0^2}{H_0}\int_{0}^{z_{\rm FRB}}\frac{f_{\rm IGM}(z)f_e(z)(1+z)}{\sqrt{\Omega_m(1+z)^3+ 1 - \Omega_m}}dz, 
\end{equation}
where $A = \frac{3c}{ 8\pi G m_p}$ and $m_p$ is the proton mass. The electron fraction is $f_e(z) = Y_H X_{e, H}(z) + \frac{1}{2} Y_{He}X_{e, He}(z)$, with hydrogen fraction
$Y_H  = 0.75$ and helium fraction $Y_{He} = 0.25$. Hydrogen and helium are completely ionized at $z<3$, which implies the ionization fractions of intergalactic hydrogen and helium $X_{e,H} = X_{e,He} = 1$. The cosmological parameters $\Omega_m$ and $\Omega_b$ are the the density of matter and the density of baryons, respectively. At present, there is no observation that can provide the evolution of the fraction of baryon in the IGM $f_{\rm IGM}$ with redshift. \citet{Shull12} provided an estimation of $f_{\rm IGM} \approx 0.83$ \cite{2022MNRAS.515L...1W}. Then, the dispersion measure-redshift relation allows FRBs to be used as cosmological probes. However, the degeneracy between $DM_{IGM}$ and $DM_{host}$ is the main obstacle for the cosmological application of FRBs. A reasonable method is to consider the probability distributions of $DM_{IGM}$ \cite{2014ApJ...780L..33M,Zhangzj2021} and $DM_{host}$ \cite{Zhanggq2020}. 

\citet{2022MNRAS.511..662H} presented the first $H_{0}$ estimation, $H_{0}$ = $62.3\pm9.1$ km/s/Mpc, using the nine then available FRBs. Employing the probability distributions of $DM_{IGM}$ \cite{Zhangzj2021} and $DM_{host}$ \cite{Zhanggq2020} from the IllustrisTNG simulation, a more compact result, $H_{0}$ = $68.81^{+4.99}_{-4.33}$ km/s/Mpc, was given by \citet{2022MNRAS.515L...1W} using eighteen localized FRBs. These two $H_{0}$ estimations seem to favor the smaller $H_{0}$ value. After that, \citet{2022MNRAS.516.4862J} show $H_{0}$ = $73^{+12}_{-8}$ km/s/Mpc employing an updated sample of 16 Australian Square Kilometre Array Pathfinder (ASKAP) FRBs 
 detected by the Commensal Real-time ASKAP Fast Transients (CRAFT) Survey and localised to their host galaxies, and 60 unlocalised FRBs from Parkes and ASKAP. Compared to previous FRB-based estimates, uncertainties in FRB energetics and $DM_{host}$ produce larger uncertainties in the inferred value of $H_{0}$. Furthermore, \citet{2022MNRAS.511..662H} performed a forecast using a realistic mock sample to demonstrate that a high-precision measurement of the expansion rate is possible without relying on other cosmological probes. Another $H_{0}$ measurement, $H_{0}$ = 70.60 $\pm$ 2.11 km/s/Mpc, was given by \citet{2022arXiv221005202L}, employing a cosmological-model independent method. Recently, \citet{2022arXiv221213433Z} provided the first statistical $H_{0}$ measurements using unlocalized FRBs. They provided two $H_{0}$ measurements, $H_{0}$ = 71.7$^{+8.8}_{-7.4}$ km/s/Mpc and $H_{0}$ = 71.5$^{+10.0}_{-8.1}$ km/s/Mpc, which were obtained from the simulation-based case and the observation-based case, respectively. They also proposed that in the next few years, a 3$\%$ precision on the random error of the Hubble constant could be achieved using thousands of FRBs.

\subsection{Tip of the Red Giant Branch}
Compared to other standard candles, the tip of the red giant branch (TRGB) offers many advantages; for example, in the K-band they are $\sim$1.6 magnitudes brighter than Cepheids. They have nearly exhausted the hydrogen in their cores and have just begun helium burning. Employing parallax methods, their brightness can be standardized. They thus can be regarded as the standard candles visible in the local universe. Instead of the Cepheid, the TRGB can be used as calibrators of SNe Ia. \citet{2019ApJ...882...34F} gave the Hubble constant result $H_{0}$ = 69.8 $\pm$ 0.8 km/s/Mpc by measuring TRGB in nine SNe Ia hosts and calibrating TRGB in the large magellanic cloud. A consistent result, $H_{0}$ = 69.6 $\pm$ 0.8 km/s/Mpc, was also given by \citet{2020ApJ...891...57F} using their revised measurement of the Large Magellanic Cloud TRGB extinction. After that, \citet{2021ApJ...919...16F} combined several recent calibrations of the TRGB method, and 
 are internally self-consistent at the 1$\%$ level. The updated TRGB calibration applied to a sample of SNe Ia from the Carnegie Supernova Project results in a value of $H_{0}$ = 69.8 $\pm$ 0.8 km/s/Mpc. 

We display $H_{0}$ measurements obtained from the recent observations in Table~\ref{tab3}, and describe the correlation between the $H_{0}$ values and the published year in Figure~\ref{fig2}. Combining Table~\ref{tab3} and Figure~\ref{fig2}, it is easy to find that the measurements of the quasar lensing are tending to the SH0ES results as a whole. The $H_{0}$ measurements obtained from Megamaser, GW and FRB are more diffuse and have a large error which both covered the SH0ES results and CMB results. It is interesting that the $H_{0}$ values measured by TRGB have small errors 
 and are between the SH0ES results and the CMB results. Collectively, these independent observations are currently unable to arbitrate the $H_{0}$ tension.

\begin{table}[H] 
\caption{$H_{0}$ measurements with the 68$\%$ confidence level derived from the recent observations.\label{tab3}}
\newcolumntype{C}{>{\centering\arraybackslash}X}
\begin{tabularx}{\textwidth}{p{2cm}p{1.5cm}p{2.5cm}p{2cm}p{1.5cm}p{2cm}}
\toprule
\textbf{Observation} & \boldmath{$H_{0}$} \textbf{(km/s/Mpc)} & \textbf{Reference} & \textbf{Observation}	& \boldmath{$H_{0}$} \textbf{(km/s/Mpc)}	& \textbf{Reference}\\
\midrule
Quasar lens & $73.3^{+1.7}_{-1.8}$ & \cite{2020MNRAS.498.1420W} & FRB & $62.3\pm9.1$ & \cite{2022MNRAS.511..662H}\\
Quasar lens & $74.0^{+1.7}_{-1.8}$	& \cite{2020AA...639A.101M} & FRB & $68.81^{+4.99}_{-4.33}$ & \cite{2022MNRAS.515L...1W}\\
Quasar lens & $74.2^{+1.6}_{-1.6}$	& \cite{2023arXiv230102656S} & FRB  & $73.0^{+12.0}_{-8.0}$ & \cite{2022MNRAS.516.4862J}\\
Megamaser  & $69.3\pm4.2$	& \cite{2018IAUS..336...86B} & FRB & 70.6 $\pm$ 2.11 & \cite{2022arXiv221005202L}\\
Megamaser  & $73.9\pm3.0$	& \cite{2020ApJ...891L...1P} & FRB  & 71.5 $^{+10.0}_{-8.1}$ & \cite{2022arXiv221213433Z}\\
GW  & $74.0^{+16.0}_{-8.0}$ & \cite{2017Natur.551...85A} & TRGB & 69.8 $\pm$ 0.8  & \cite{2019ApJ...882...34F} \\
GW + EM & $70.3^{+5.3}_{-5.0}$ & \cite{2019NatAs...3..940H} & TRGB  & 69.6 $\pm$ 0.8 & \cite{2020ApJ...891...57F} \\
GW  & $68.0^{+12.0}_{-7.0}$ & \cite{2021arXiv211103604T} & TRGB  & 69.8 $\pm$ 0.8  & \cite{2021ApJ...919...16F}   \\
GW  & $67.0^{+6.3}_{-3.8}$ & \cite{2022arXiv220303643M} &   &   &    \\
\bottomrule
\end{tabularx}
\end{table}
\vspace{-12pt}

\begin{figure}[H]
\includegraphics[width=7 cm]{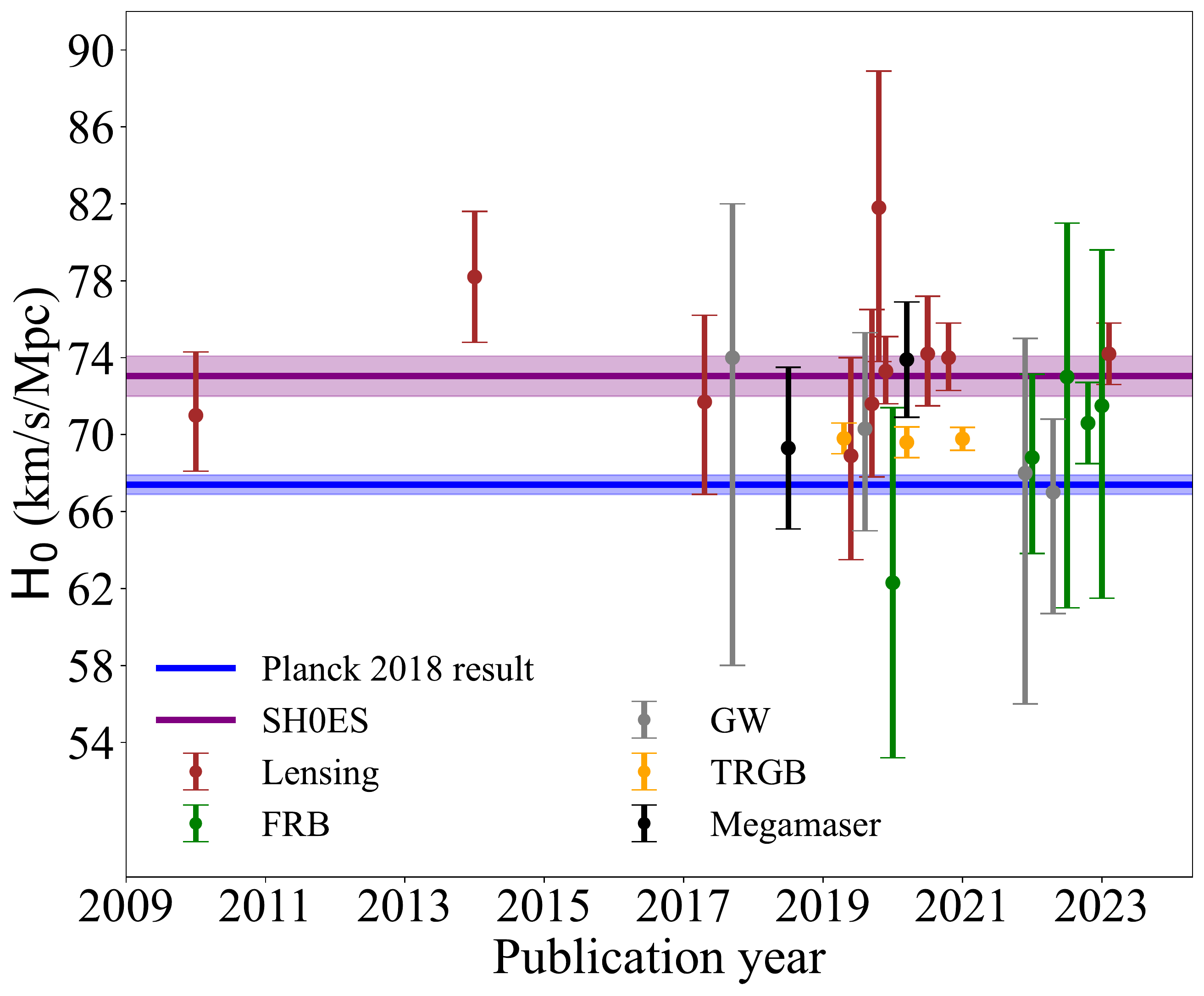}
\caption{$H_{0}$ measurements in the late-time universe derived from other independent observations which include quasar lensing \cite{2010AA...524A..94S,2014ApJ...788L..35S,2017MNRAS.465.4895W,2019Sci...365.1134J,2019MNRAS.490.1743C,2019MNRAS.484.4726B,2020MNRAS.498.1440R,2020AA...639A.101M,2020MNRAS.498.1420W,2023arXiv230102656S}, FRB \cite{2022MNRAS.511..662H,2022MNRAS.515L...1W,2022arXiv221213433Z}, GW \cite{2017Natur.551...85A,2019NatAs...3..940H,2021arXiv211103604T,2022arXiv220303643M}, TRGB \cite{2019ApJ...882...34F,2020ApJ...891...57F,2021ApJ...919...16F} and Megamaser \cite{2018IAUS..336...86B}. The purple
and blue regions correspond to the results of SH0ES \cite{2022ApJ...934L...7R} and Planck collaborations \cite{2020A&A...641A...6P}, respectively.
\label{fig2}}
\end{figure}   
\unskip

\section{Solutions for the \boldmath{$H_{0}$} Tension}\label{sec4}
The case for an observational discrepancy between the early and late universe appears strong, is hard to dismiss, and merits an explanation. The analyses of possible systematics did not lead us to the cause of the Hubble tension \cite{2017A&A...607A..95P,2018ApJ...867..108J,2019ApJ...876...85R,2019MNRAS.484L..64S,2020A&A...641A...1P,2020A&A...641A...7P,2020A&A...644A.176R,2022MNRAS.514.4620D,2022ApJ...934L...7R}. The arbitration of $H_{0}$ given by otherwise independent observations also did not produce consistent results in favor of one side \cite{2017Natur.551...85A,2018IAUS..336...86B,2019NatAs...3..940H,2020AA...639A.101M,2020MNRAS.498.1420W,2022MNRAS.511..662H,2022MNRAS.515L...1W,2021ApJ...919...16F}. Hence, many researchers prefer to believe that the Hubble tension may be caused by new physics beyond the $\Lambda$CDM model \cite{2020PhRvD.102b3518V}. Until now, a large number of theoretical solutions have been proposed to solve or relieve the $H_{0}$ tension. The detailed information can be found from these references \cite{2021CQGra..38o3001D,2021A&ARv..29....9S,2022NewAR..9501659P}. 
\subsection{Classification of Solutions to $H_{0}$ Tension}\label{sec41}
In the review article for $H_{0}$ tension, \citet{2021CQGra..38o3001D} gave a detailed classification for the solutions of the $H_{0}$ tension. They divided all schemes proposed for solving the $H_{0}$ tension into 11 major categories with 123 subcategories, and some subcategories include several different schemes. The detailed classification results are as follows:
\begin{enumerate}
\item [(a)] Early dark energy \cite{2021PhRvD.103j3506S,2022arXiv221104492K,2022arXiv221016296H}:

(1) Anharmonic oscillations \cite{2019PhRvL.122v1301P}; \\
(2) Ultra-light axions \cite{2019IJMPD..2844017K,2020PhRvD.102d3507H,2020PhLB..81035791L,2021PhRvD.103d3529C,2021JCAP...05..072D,2021PhRvD.103f3539H,2021PhRvD.103f3502M}; \\
(2.1) Dissipative axion \cite{2020PhRvD.101h3537B};\\
(2.2) Axion interacting with a dilaton \cite{2019PhLB..79734830A}; \\
(3) Power-law potential \cite{2020JCAP...08..013C}; \\
(4) Rock 'n' roll \cite{2019arXiv190401016A}; \\
(5) New early dark energy \cite{2020PhRvD.102f3527N,2021PhRvD.103d1303N};\\
(6) Chain early dark energy \cite{2021PhRvD.104h3533F};\\
(7) Anti-de Sitter phase \cite{2020PhRvD.101h3507Y,2022arXiv221204429O};\\
(8) Gradusted dark energy \cite{2020PhRvD.101f3528A}; \\
(9) Acoustic dark energy \cite{2019PhRvD.100f3542L,2022EPJC...82...78Y}; \\
(9.1) Exponential acoustic dark energy \cite{2022EPJC...82...78Y}; \\
(10) Early dark energy in $\alpha$ attractors \cite{2020PhRvD.102b3529B}. 
\item [(b)] Late dark energy \cite{2017NatAs...1..627Z,2021Univ....8...22B,2022arXiv220111623H}:

(1) $w$CDM model \cite{2019Symm...11..986M,2020PhRvD.101l3516A,2020PhRvD.102b3518V,2021JCAP...01..006D};\\
(2) $w_{0}w_{a}$CDM or CPL parameterization \cite{2019Symm...11..986M,2019Univ....5..219Y,2020PhRvD.101l3516A};\\
(3) Dark energy in extended parameter spaces \cite{2020JCAP...01..013D};\\
(4) Dynamical dark energy parameterizations with two free parameters \cite{2018PhRvD..98h3501V,2019PhRvD.100d3535D};\\
(5) Dynamical dark energy parameterizations with one free parameter \cite{2019PhRvD..99d3543Y};\\
(6) Matastable dark energy \cite{2019ApJ...887..153L,2020JCAP...04..029S,2020PhRvD.102f3503Y};\\ 
(7) Phantom crossing \cite{2021Entrp..23..404D};\\
(8) Late dark energy transition \cite{2020PhRvD.101j3517B,2021PhRvD.103h3517A};\\
(9) Running vacuum model \cite{2021EL....13419001S,2021EPJST.230.2077M};\\
(10) Transitional dark energy model \cite{2019JCAP...12..035K};\\
(11) Negative dark energy \cite{2020GReGr..52...15D};\\
(12) Bulk viscous models \cite{2019PhRvD.100j3518Y,2020PhRvD.102l3501E,2021EPJC...81..403D};\\
(13) Holographic dark energy \cite{2019JCAP...02..054G,2020MNRAS.491L...6V,2020PhRvD.102l1302D,2021CQGra..38q7001C};\\
(13.1) Tsallis holographic dark energy \cite{2021EPJP..136..543D};\\
(14) Swampland conjectures \cite{2019PhLB..793..126O,2019PhLB..79734907O,2020PhRvD.101h3532A,2021PhRvD.103d3523A,2021PhRvD.103h1305B};\\
(15) Late time transitions in the quintessence field \cite{2018ApJ...868...20M,2019PDU....2600385D};\\
(16) Phantom braneworld dark energy \cite{2021ApJ...923..212B};\\
(17) Frame-dependent dark energy \cite{2019PhRvD.100l3503A};\\
(18) Chameleon dark energy \cite{2021PhRvD.103l1302C,2021PhRvD.103l1302C,2022PhRvD.105f3535K}.
\item [(c)] Dark energy models with degrees of freedom and their extensions: 

(1) Phenomenologically emergent dark energy \cite{2019ApJ...883L...3L,2020MNRAS.497.1590H,2021PDU....3100762Y};\\
(1.1) Generalized emergent dark energy \cite{2020ApJ...902...58L,2021PhRvD.104f3521Y};\\
(1.2) Modified emergent dark energy \cite{2020arXiv200809098B};\\
(2) Vacuum metamorphosis \cite{2018PhRvD..97d3528D,2020PDU....3000733D,2021PhRvD.103l3527D};\\
(2.1) Elaborated vacuum metamorphosis \cite{2018PhRvD..97d3528D,2020PDU....3000733D,2021PhRvD.103l3527D}.
\item [(d)] Models with extra relativistic degrees of freedom: 

(1) Sterile neutrinos \cite{2019PhRvD.100b3505C,2020JCAP...09..051G};\\
(2) Neutrino asymmetries \cite{2017EPJC...77..590B};\\
(3) Thermal axions \cite{2018JCAP...11..014D,2020PhR...870....1D};\\  
(4) Decaying dark matter \cite{2020JCAP...01..045X,2020JCAP...06..005B,2020MNRAS.497.1757H,2020JCAP...07..026P,2021PhRvD.103c5025A,2022MNRAS.516.4373D};\\
(4.1) Self-interacting dark matter \cite{2020PhRvD.102d3024H,2022epsc.confE.115J};\\
(4.2) Two-body dark matter decays \cite{2019PhRvD..99l1302V,2021PhRvD.103d3014C};\\
(4.3) Light gravitino scenarios \cite{2020PhLB..80535408C,2020PhRvD.102k5005G,2021PhLB..81236008A};\\
(4.4) Decaying 'Z' \cite{2019arXiv190102010E};\\
(4.5) Dynamical dark matter \cite{2020PhRvD.101c5031D};\\
(4.6) Degenerate decaying fermion dark matter \cite{2020PhRvD.101g5031C};\\
(5) Neutrino--dark matter interactions \cite{2018PhRvD..97d3513D,2019JCAP...08..014S,2021JCAP...03..066M};\\
(5.1) Neutrino--Majoron interactions \cite{2021EPJC...81...28A,2021EPJC...81..515E,2021PhRvD.103l3007H};\\ 
(5.2) Feebly interacting massive particles (FIMPs) decay into neutrinos \cite{2021PhRvD.104c5006B};\\  
(6) Interacting dark radiation \cite{2021PhRvD.104c5006B};\\
(7) Coupled DM--dark radiation scenarios \cite{2019JCAP...10..055A,2021JCAP...02..019B};\\
(8) Cannibal dark matter \cite{2018PhRvD..98h3517B};\\
(9) Decaying ultralight scalar \cite{2018PhRvD..98h3517B,2018JCAP...05..067F};\\
(10) Ultralight dark photon \cite{2019PhRvD.100l3525A,2019PhRvD.100f3541F};\\
(11) Primordial black holes \cite{2020ARNPS..70..355C,2021PhRvL.126d1101F,2021JPhG...48d3001G};\\
(12) Unparticles \cite{2021PhRvD.103l1303A}.
\item [(e)] Models with extra interactions: 

(1) Interacting dark energy (IDE) \cite{2022arXiv220705955Y};\\
(1.1) Interacting vacuum energy \cite{2019EPJC...79..576K,2020JCAP...04..008Y,2020PDU....3000666D,2021JCAP...07..005G,2022MNRAS.514.1433W};\\
(1.2) Coupled scalar field \cite{2020PhRvD.101l3513G};\\ 
(1.3) IDE with a constant DE equation of state \cite{2018JCAP...09..019Y,2020PhRvD.101f3502D,2019PhRvD..99h3509B,2020MNRAS.493.3114P};\\
(1.4) IDE with variable DE equation of state \cite{2019PhRvD.100j3520P,2021PhRvD.103l3527D};\\
(1.5) Interacting vacuum scenario and IDE with variable coupling \cite{2019PhRvD..99h3509B,2020PhRvD.102b3535Y};\\
(1.6) IDE with sign-changing interaction \cite{2019PhRvD.100h3539P};\\
(1.7) Anisotropic stress in IDE \cite{2019MNRAS.482.1858Y};\\
(1.8) Interaction in the anisotropic universe \cite{2020arXiv200103775A};\\
(1.9) Metastable interacting dark energy \cite{2019ApJ...887..153L,2020PhRvD.102f3503Y};\\
(1.10) Quantum field cosmology \cite{2019NuPhB.940..312B,2021JCAP...07..005G};\\
(1.11) Interacting quintom dark energy \cite{2021ChPhC..45a5108P};\\
(2) Interacting dark matter \cite{2020arXiv201207494J,2022PDU....3801131H};\\
(2.1) DM--photon coupling \cite{2018JCAP...10..009S,2020MPLA...3550358Y};\\
(2.2) DM--baryon coupling \cite{2018Natur.555...71B,2018PhRvD..98b3013S};\\
(3) DE--baryon coupling \cite{2020JCAP...08..020B,2020MNRAS.493.1139V};\\
(4) Interacting neutrinos \cite{2019PhRvL.123s1102B,2020JCAP...11..003H,2021PhRvD.103a5004L};\\ 
(4.1) Self-interacting neutrinos: \cite{2019PhRvL.123s1102B,2020JCAP...11..003H,2020PhRvD.101l3505K,2021PhRvD.104f3523B,2022arXiv220301955B};\\
(4.2) Self-interacting sterile neutrino model \cite{2020JCAP...12..029A};\\
(4.3) Dark neutrino interactions \cite{2018PhRvD..97f3529G,2020PhRvD.102l3544G}.
\item [(f)] Unified cosmologies: 

(1) Generalized Chaplygin gas model \cite{2019JCAP...11..044Y};\\
(2) A new unified model \cite{2019MNRAS.490.2071Y};\\
(3) $\Lambda$(t)CDM model \cite{2021arXiv210210123B};\\
(4) $\Lambda$-gravity \cite{2019EPJC...79..568G,2021EPJP..136..235G}.
\item [(g)] Modified gravity \cite{2019IJMPD..2830012Q}: 

(1) f($\mathcal{R}$) gravity theory; \cite{2020PhRvD.101j3505D,2021NuPhB.96615377O,2021EPJC...81..482W,2022arXiv221116737S}; \\
(2) f($\mathcal{T}$) gravity theory \cite{2018JCAP...05..052N,2020PhRvD.101l1301Y,2020PhRvD.102f3530W,2022arXiv220506252A}; \\ 
(3) f($\mathcal{T, B}$) gravity theory \cite{2020CQGra..37p5002E,2021Univ....7..150P}; \\
(4) f($\mathcal{Q}$) gravity theory \cite{2022PDU....3500980A,2022arXiv220804723K}; \\
(5) Jordan--Brans--Dicke (JBD) gravity \cite{2022PhRvD.105d3522J};\\ 
(5.1) Brans and Dicke-$\Lambda$CDM \cite{2019ApJ...886L...6S,2020CQGra..37x5003P}; \\ 
(6) Scalar-tensor theories of gravity \cite{2020JCAP...10..044B,2020JCAP...11..024B}; \\ 
(6.1) Early modified gravity \cite{2021PhRvD.103b3530A,2021PhRvD.103d3528B}; \\ 
(6.2) Screened fifth forces \cite{2020PhRvD.101l9901D,2020PhRvD.102b3007D}; \\ 
(7) \''{U}ber-gravity \cite{2018PDU....21...21K,2019PhRvD..99j3526K}; \\ 
(8) Galileon gravity \cite{2020PhRvD.102b3523Z,2021JCAP...03..032H}; \\
(9) Nonlocal gravity \cite{2018JCAP...03..002B,2020JCAP...04..010B}; \\
(10) Unimodular gravity \cite{2021PDU....3200807L}; \\
(11) Scale-dependent scenario of gravity \cite{2021JCAP...06..019A}; \\
(12) VCDM \cite{2021PhLB..81636201D,2022arXiv221213561G}. 
\item [(h)] Inflationary models \cite{2022PhRvD.106l3524R}:

(1) Exponential inflation \cite{2017JCAP...03..020D,2017EPJC...77..882G}; \\
(2) Reconstructed primordial power spectrum \cite{2020JCAP...09..055K,2020ApJ...897..166L}; \\
(3) Lorentzian quintessential inflation \cite{2021JCAP...07..007A}; \\
(4) Harrison--Zel'dovich spectrum \cite{2018PhRvD..98f3508D}. 
\item [(i)] Modified recombination history \cite{2018arXiv181103624C}:

(1) Effective electron rest mass \cite{2018MNRAS.474.1850H,2020MNRAS.493.3255H}; \\
(2) Time varying electron mass \cite{2021PhRvD.103h3507S}; \\ 
(3) Axi--Higgs model \cite{2021JCAP...08..057F}; \\
(4) Primordial magnetic fields \cite{2019PhRvL.123b1301J,2020PhRvL.125r1302J}. 
\item [(j)] Physics of the critical phenomena: 

(1) Double-$\Lambda$CDM \cite{2020PhRvD.101l3521B}; \\
(2) Ginzburg--Landau theory of phase transition \cite{2019PhRvD..99h3509B}; \\
(3) Critically emergent dark energy \cite{2021JCAP...06..003B}.
\item [(k)] Alternative proposals: 

(1) Local inhomogeneity \cite{2019PTEP.2019g3E01K,2022arXiv220412180Y}; \\ 
(2) Bianchi type I spacetime \cite{2019PhRvD.100b3532A}; \\
(3) Scaling solutions \cite{2018PhRvD..97j3529B,2018ApJ...865L...4M,2020CQGra..37p4001H}; \\
(4) CMB monopole temperature $T_{0}$ \cite{2020PhRvD.102f3515I}; \\
(4.1) Open and hotter universe \cite{2020EPJC...80..936B,2021PhRvD.103h1304B}; \\
(5) Super-$\Lambda$CDM \cite{2020PDU....2800539A}; \\ 
(6) Heisenberg uncertainty \cite{2020FoPh...50..893C} \\
(7) Diffusion \cite{2021GReGr..53....7P}; \\
(8) Casimir cosmology \cite{2021MNRAS.507.3473B}; \\
(9) Surface forces \cite{2020IJMPD..2950115O}; \\
(10) Milne model \cite{2020IJMPD..2943025V,1935rgws.book.....M}; \\ 
(11) Running Hubble tension \cite{2021PhRvD.103j3509K,2021ApJ...912..150D}; \\
(12) Rapid transition in the effective gravitational constant \cite{2021PhRvD.104b1303M}; \\
(13) Causal horizons \cite{2021MNRAS.504.5840F,2020MNRAS.494.2766G}; \\
(14) Milgromian dynamics \cite{2020MNRAS.499.2845H}; \\
(15) Charged dark matter \cite{2020IJMPD..2943010B,2021PhRvD.103j3505J}. 
\end{enumerate}

It can be observed that \citet{2021CQGra..38o3001D}'s classification of schemes to alleviate the $H_{0}$ tension is very detailed. With help of this classification, we can quickly find out the solutions and corresponding articles we need. Recently, \citet{2022NewAR..9501659P} updated and optimized the classification scheme based on the latest research work. The new classification scheme is more concise than before. They divided all solutions into 5 major categories, each of which contained several sub-categories, for a total of 19 sub-categories. The detailed classification is as follows:
\begin{enumerate} 
\item [(i)] Late time deformations of the Hubble expansion rate H(z):  

(1) Phantom dark energy \cite{2022PhRvD.106j3508C};\\
(2) Running vacuum model \cite{2021arXiv210212758S};\\
(3) Phenomenologically emergent dark energy \cite{2020JCAP...06..062P,2020IJMPD..2950088L,2022PhRvD.105j3511D};\\ 
(4) Vacuum phase transition \cite{2022ApJ...940..121M};\\
(5) Phase transition in dark energy \cite{2021PhRvD.104f3506M}.
\item [(ii)] Deformations of the Hubble expansion rate H(z) with additional interactions/degrees of freedom:

(1) Interacting dark energy \cite{2018PhRvD..98l3527Y,2021JCAP...10..004C,2021JCAP...12..036G,2021PhRvD.104f3529N,2022JHEAp..36..141C,2022PhRvD.106b3530G};\\
(2) Decaying dark matter \cite{2022PhRvD.105j3512A}.
\item [(iii)] Deformations of the Hubble expansion rate H(z) with inhomogeneous or anisotropic modifications:

(1) Chameleon dark energy \cite{2021PhRvD.104f3023V,2022PhRvD.105b4052B};\\
(2) Cosmic voids \cite{2020PhLB..80335303L,2022arXiv221207438C};\\
(3) Inhomogeneous causal horizons \cite{2021MNRAS.504.5840F}.
\item [(iv)] Late time modifications: Transition or recalibration of the SNe Ia absolute luminosity:

(1) Gravity and evolution of the SNe Ia intrinsic luminosity \cite{2020PhRvD.102b3520K};\\
(2) Transition of the SNe Ia absolute magnitude $M$ at a redshift z $\simeq$ 0.01 \cite{2021MNRAS.504.3956A,2022Univ....8..502P,2022PhRvD.106d3528P};\\
(3) Late (low-redshift) $w - M$ phantom transition \cite{2020ApJ...894...54D,2021PhRvD.104l3511P,2022ApJ...935...58M}.
\item [(v)] Early time modifications of sound horizon:

(1) Early dark energy \cite{2020PhRvD.101f3523S,2020PhRvL.124p1301S,2021ApJ...915..132G,2021PhRvD.104l3550P,2021PhRvD.103l3501S,2021arXiv210713391Y,2021PhRvD.104j3524J,2022PhLB..83537555N,2022PhRvD.106d3526S};\\ 
(2) Dark radiation \cite{2021EPJC...81..954F,2021PhRvD.104f3019S,2022JCAP...04..042A,2022PhRvD.105i5008G,2022PhRvD.105l3516A,2022JCAP...05..014G};\\
(3) Neutrino self-interactions \cite{2020PhRvD.102k5008B,2021JCAP...03..084C,2021PhRvD.103a5004L,2022JCAP...10..011M};\\
(4) Large primordial non-Gaussianities \cite{2020PDU....2800539A};\\
(5) Heisenberg's uncertainty principle \cite{2020FoPh...50..893C};\\
(6) Early modified gravity \cite{2020JCAP...10..044B,2020JCAP...11..024B,2021PhRvD.103b3530A}.
\end{enumerate}

Of course, the above classifications may not completely cover all the solutions to relive $H_{0}$ tension. The proposal of new schemes never stops, such as the Weyl invariant gravity \cite{2022JCAP...04..048S}, Horndeski gravity \cite{2022PhRvD.106l4051P}, $\Lambda_{S}$CDM model \cite{2021PhRvD.104l3512A,2022arXiv221105742A}, Early Integrated Sachs--Wolfe (eISW) effect \cite{2021PhRvD.104f3524V}, realistic model of dark atoms \cite{2022PhRvD.105i5005B}, information dark energy \cite{2022Entrp..24..385G}, U(1)$_{L_\mu - L_\tau}$ model with Majoron \cite{2021PTEP.2021j3B05A}, etc.

Classification is a summary of previous research work on relieving the $H_{0}$ tension, which helps to find out the physical origin that causes the discrepancy in $H_{0}$ measurements. For the classification of solutions to the $H_{0}$ tension schemes, not all schemes fall perfectly into one of these categories \cite{2022PhR...984....1S}. At present, the proposed models and theories are usually divided into three categories: early-time model, late-time model and modified gravity. The boundary between the early-time models and late-time models is the recombination redshift ($z \sim 1100$). Detailed introduction of each scheme has been made in previous $H_{0}$ review literature \cite{2021CQGra..38o3001D} and \citet{2022NewAR..9501659P}. We will not repeat it here. In addition to categorizing so many solutions, there are several works that provide comparative analysis of solutions for $H_{0}$ tension, and discussion on the $H_{0}$ tension \cite{2019PhRvD.100d3537D,2020ApJ...897..166L,2020auhc.confE..22P,2021ApJ...912L...1A,2021PhRvD.104f3535T,2021CmPhy...4..123J,2022PhRvD.105b1301C,2022PhRvD.106f3519C,2022PhRvD.106j3517E,2022PhRvD.106b3011W,2022PhR...984....1S,2023arXiv230103883K}. \citet{2022PhR...984....1S} organized $H_{0}$ Olympic-like games for the relative success of seventeen models which have been proposed to resolve the $H_{0}$ tension, and gave a ranking. Finally, the early dark energy model, new early dark energy model, early modified gravity model and varying $me$+$\Omega_{K}$ model are the most successful of the models studied in the $H_{0}$ Olympic-like games.

According to the research and discussions on 
 solving the $H_{0}$ tension, we tend to divide all of the solutions into two categories: one proposes a new cosmological model first, and then combines the existing early-time (for example, CMB) or late-time (for example, Cepheid) observational data-sets to constrain the cosmological parameters (including $H_{0}$). A higher $H_{0}$ value which is consistent with the SH0ES results can be given by utilizing the recent available observations. This category is the most common used to relieve the $H_{0}$ tension. We define such schemes as the sequential scheme. There is one thing to note here. Considering that 
  the new model needs to introduce additional parameters, and the increase in the volume of the parameter space will make the final $H_{0}$ result incompact, this would amplify the extent to which the new model mitigates the $H_{0}$ tension. Moreover, it is diffcult to estimate the contribution of additional parameters to the degree of mitigation of the $H_{0}$ tension. The other is to use the $\Lambda$CDM model or model-independent methods, combined with the existing late-time observations center to find the $H_{0}$ singular behaviors which can be used to resolve or alleviate the $H_{0}$ tension. Compared to the sequential scheme, such a scheme should be called a reverse-order scheme. These $H_{0}$ singular behaviors might hint to 
   new physics beyond the $\Lambda$CDM model, and still require new cosmological models to explain them. The proposal of the new cosmological model no longer directly alleviates the $H_{0}$ tension, but explains the $H_{0}$ singular behaviors. It seems that there is no need to worry about the problem of increasing the parameter space with the additional parameters. The former scheme has been elaborated in many review articles \cite{2021CQGra..38o3001D,2022NewAR..9501659P} and will not be repeated here. We will revisit the latter scenario, which might hint at new physics beyond the $\Lambda$CDM model, in the next section.

\section{Evidence of New Physics Beyond the $\Lambda$CDM} \label{sec5}
\begin{figure}[H]
\includegraphics[width=7 cm]{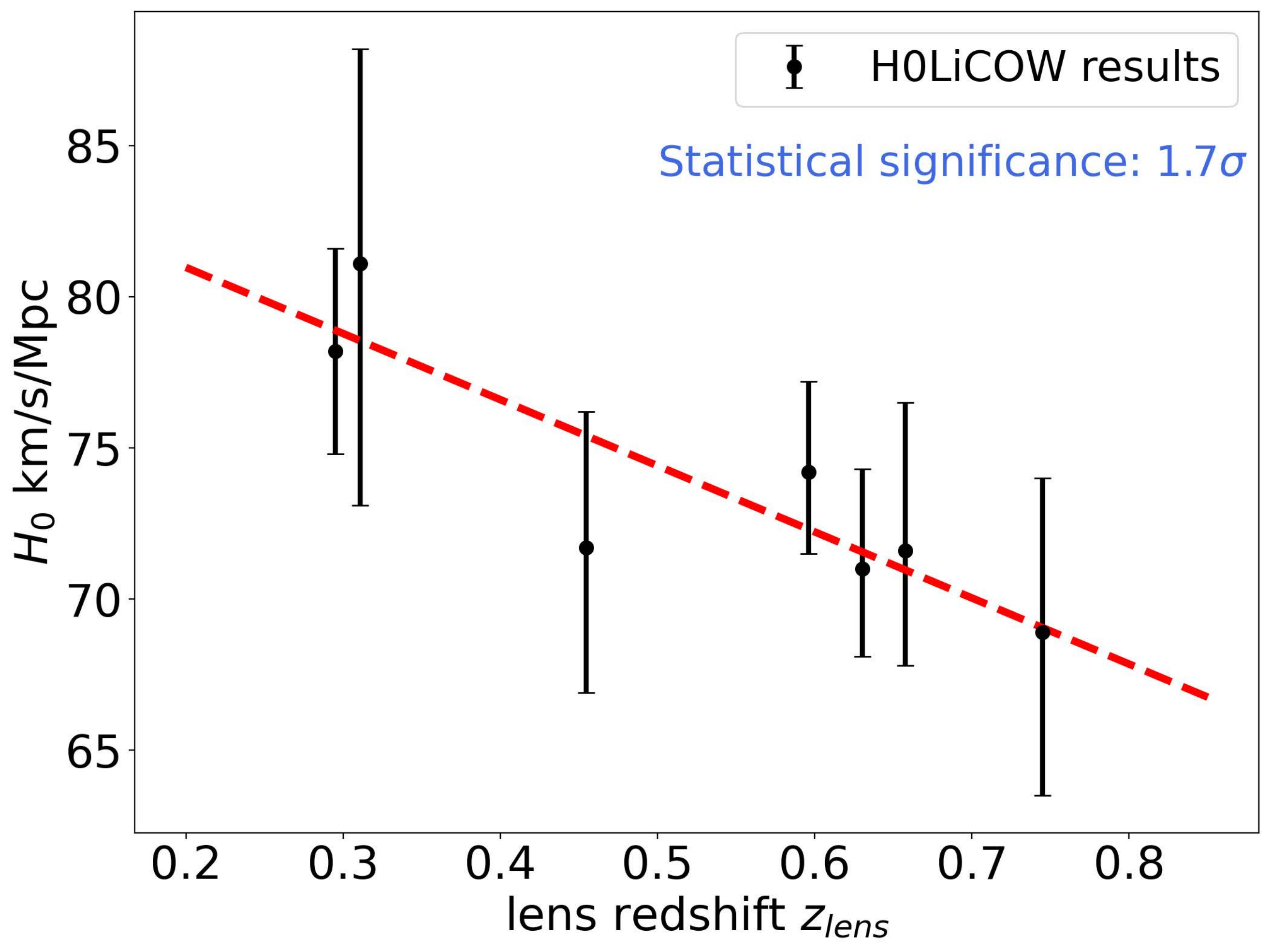}
\caption{$H_{0}$ measurements from H0LiCOW. The trend of smaller $H_{0}$ value with increasing lens redshift has significance levels of 1.7$\sigma$. (Source: Figure~5 in \citet{2020AA...639A.101M}.)
\label{fig3}}
\end{figure}

According to previous analyses of the quasar lensing, \citet{2020MNRAS.498.1420W} found that the inferred value of $H_{0}$, which was estimated using the strongly-lensed quasar time delay (H0LiCOW), decreases with the lens redshift. The H0LiCOW $H_{0}$ descending trend is of low statistical significance at 1.9$\sigma$. Adding a new H0LiCOW $H_{0}$ result \cite{2020MNRAS.494.6072S} (DES J0408-5354) reduces the statistical significance to 1.7$\sigma$ \cite{2020AA...639A.101M}, as shown in Figure~\ref{fig3}. After that, the TDCOSMO IV re-analysis lowers the $H_{0}$ estimates and increases the error bar \cite{2020A&A...643A.165B}. Hence, the significance of the $H_{0}$ descending trend may be lower, and it is not clear that the $H_{0}$ descending trend is not a systematically driven-by-analysis choice 
 in its rather complicated pipeline \cite{2020A&A...643A.165B}. Even so, this still provides a new diagnostic for the $H_{0}$ tension.

\begin{figure}[H]
\includegraphics[width=6.7cm]{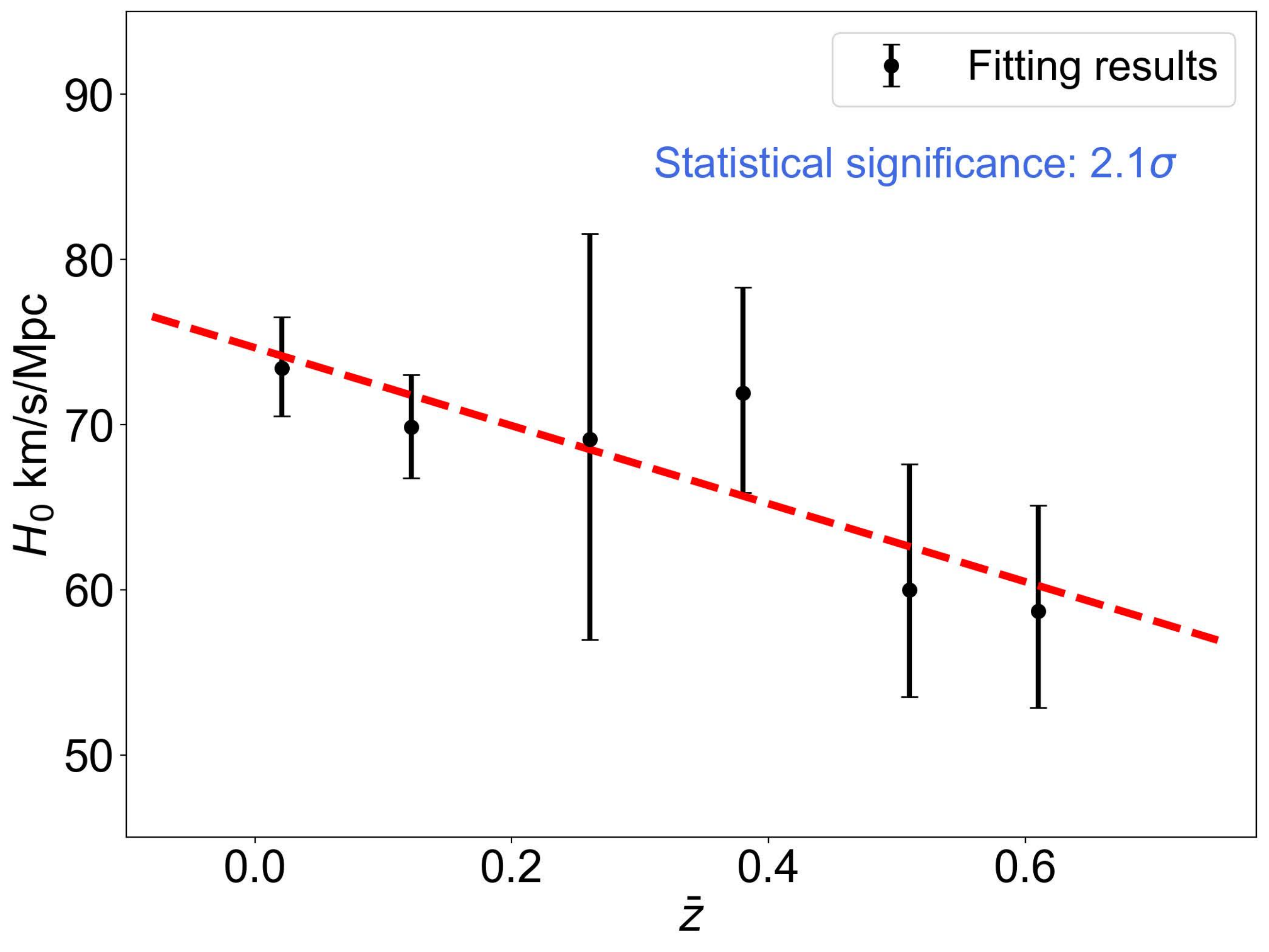}
\includegraphics[width=6.9cm]{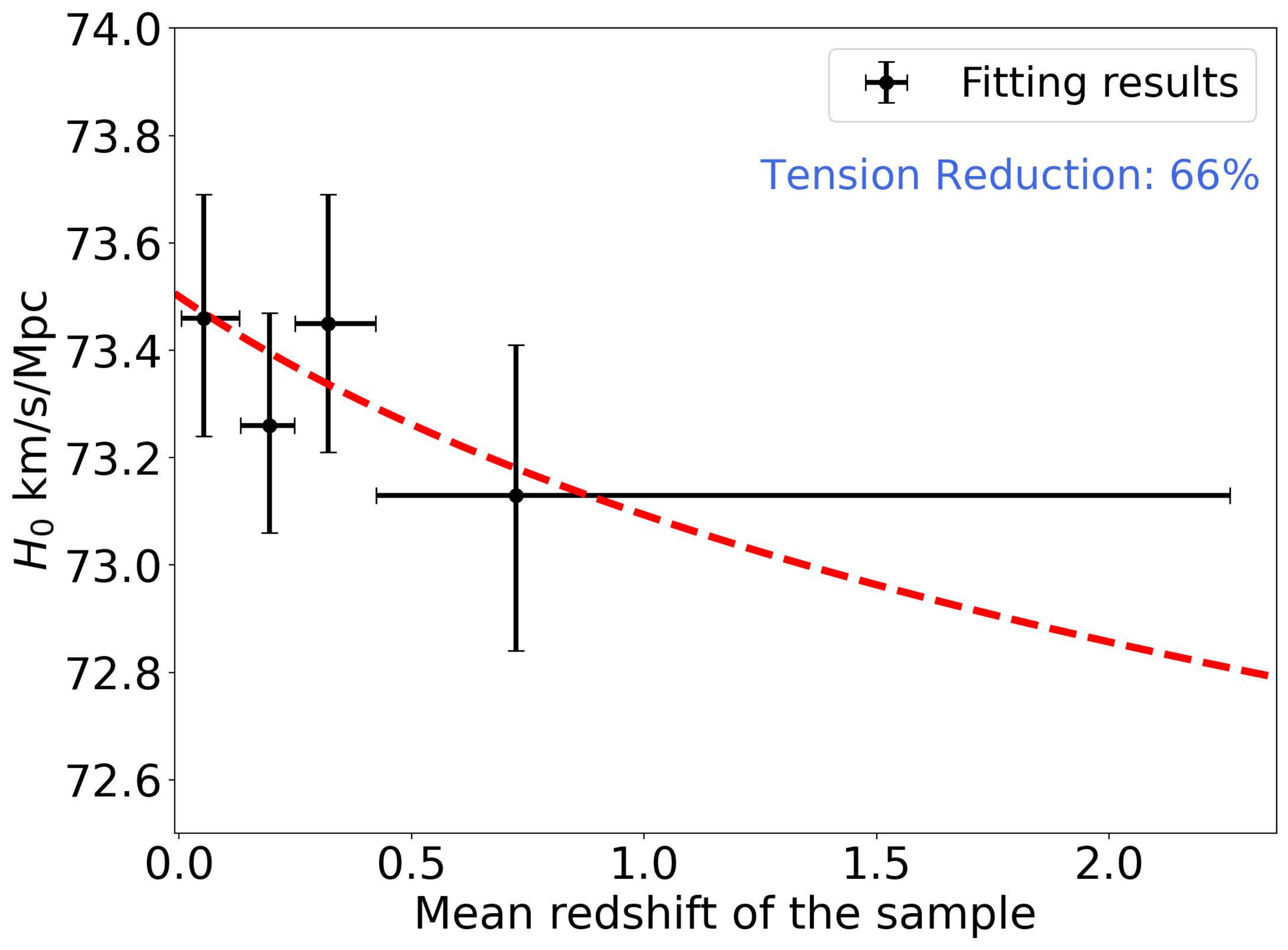}
\caption{A similar $H_{0}$ descending trend obtained from the combined sample (including megamasers, CCs, SNe Ia and BAOs) and Pantheon sample. Left panel shows the result obtained from the combined sample. The statistical significance of descending trend is 2.1$\sigma$. Right panel shows the result of Pantheon sample which can reduce the $H_{0}$ tension by 66$\%$. (Source: left panel, Figure~2 in \citet{2020PhRvD.102j3525K}; right panel, Figure~5 in \citet{2021ApJ...912..150D}.)  
\label{fig4}}
\end{figure}

Motivated by the H0LiCOW results \cite{2020MNRAS.498.1420W}, \citet{2020PhRvD.102j3525K} focused their attention mainly on late-time observations. They estimated the cosmological parameters in different redshift ranges by binning the observational data-sets (z $<$ 0.7) comprising megamasers, CCs, SNe Ia and BAOs according their redshifts. The total data-set is divided into six parts. Then, they constrained the $H_{0}$ value for each of the bin, and using ($\bar{z}$, $H_{0}$) present the final result. The form of $\bar{z}$ is written as \cite{2020PhRvD.102j3525K}
\begin{linenomath}
\begin{equation}\label{eq:z}
\bar{z} = \frac{\sum_{n}^{N_i}z_n(\sigma_{n})^{-2}}{\sum_{n}^{N_i}(\sigma_{n})^{-2}}, 
\end{equation}
\end{linenomath}
where $\sigma_{k}$ denotes the error in the observable at redshift $z_{n}$. Finally, they found a similar $H_{0}$ descending trend in the $\Lambda$CDM model with a low statistical significance (2.1$\sigma$), as shown in the left panel of Figure~\ref{fig4}. This result obtained by using the observation data completely independent of H0LiCOW is consistent with the H0LiCOW results to a certain extent. 
\unskip
\vspace{0.8em}

Based on the flat $\Lambda$CDM model and flat $w_{0}w_{a}$CDM model, a similar $H_{0}$ descending trend is also found by \citet{2021ApJ...912..150D} using the Pantheon sample only. In this analysis, they set the absolute magnitude of SNe Ia so that $H_{0}$ = 73.5 km/s/Mpc, and they fix fiducial values for $\Omega_{0m}^{\Lambda CDM}$ = 0.298 and $\Omega_{0m}^{\Lambda CDM}$ = 0.308. The $g(z)$ function was used to describe the behavior of the $H_{0}$ descending. Its form is as follows: 
\begin{linenomath}
\begin{equation}\label{eq:gz}
g(z) = H_{0}(z) = \frac{\tilde{H_{0}}}{(1+z)^{\alpha}}, 
\end{equation}
\end{linenomath}
where $\tilde{H_{0}}$ and $\alpha$ are free parameters, and $\alpha$ indicates the evolutionary trend. The $g(z)$ function is the standard for characterizing the evolution of many astrophysical sources and is widely used for GRBs and quasars \cite{2000ApJ...543..722L,2011ApJ...743..104S,2020ApJ...904...97D}. In addition, they also considered four kinds of binning methods: 3 bins, 4 bins, 20 bins and 40 bins. Finally, they reduced the $H_{0}$ tension in the range of (54$\%$, 72$\%$) for both cosmological models and pointed out that the $H_{0}$ descending trend is independent of the cosmological model and number of bins. A more detailed result can be found from Table~1 in \citet{2021ApJ...912..150D}. 

Here, we demonstrate their fitting results obtained from the 4 bins method in the flat $\Lambda$CDM model. The results of $\tilde{H_{0}}$ and $\alpha$ are 73.493 $\pm$ 0.144 km/s/Mpc and 0.008 $\pm$ 0.006, respectively. Adopting the $g(z)$ function, they obtained $H_{0}(z = 1100)$ = 69.271 $\pm$ 2.815 km/s/Mpc, which can be used to reduce the $H_{0}$ tension effectively, as shown in the right panel of Figure~\ref{fig4}. The reduction of the $H_{0}$ tension is 66$\%$. A similar descending trend (i.e., $H_{0}$ decreasing as $z_{min}$ increases) was also found by \citet{2022A&A...668A..34H} and \citet{2022PhRvD.106d1301O} from the Pantheon sample. The constraints of $H_{0}(z_{min})$ are obtained by using the SNe Ia data larger than $z_{min}$.

\unskip 
\vspace{0.8em}

After that, \citet{2022Galax..10...24D} made a further analysis for the $H_{0}$ evolution, which was described by Equation (\ref{eq:gz}) combining the Pantheon sample and BAO data. The final results demonstrate that a descending trend with $\alpha \sim 10^{-2}$ is still visible in the combined sample. The $\alpha$ coefficient reaches zero in 2.0 $\sigma$ and 5.8 $\sigma$ for the $\Lambda$CDM model and $w_{0}w_{a}$CDM model, respectively. In addition, \citet{2022arXiv220611447C} performed an independent investigation in the $\Lambda$CDM model, adopting the composite sample consisting of H(z), SNe Ia and quasars. In the end, they confirmed that the observations exhibit an increasing $\Omega_{m}$ (decreasing $H_{0}$) trend with an increasing bin redshift and that such behaviour can arise randomly within the flat $\Lambda$CDM model with a lower probability $p$ = 0.0021 (3.1 $\sigma$). 
 
At the same time, \citet{2021PhRvD.104l3511P} proposed a physically motivated new degree of freedom in the Cepheid calibrator analysis, allowing for a transition in one of the Cepheid modeling parameters $R_{w}$ (Cepheid Wesenheit color-luminosity parameter) or $M_{w}$ (Cepheid Wesenheit H-band absolute magnitude). This is mildly favored by the observational data and can yield a lower $H_{0}$ value. Their results may imply the presence of a fundamental physics transition taking place at a time more recent than 100 $Myrs$ ago. The transition magnitude is consistent with the magnitude required for the resolution of the $H_{0}$ tension in the context of a fundamental gravitational transition occurring by a sudden increase in the strength of the gravitational interactions $G_{eff}$ by about 10$\%$ \cite{2021PhRvD.104b1303M} at a redshift $z_{t}$ $\lesssim$ 0.01. Such a transition would abruptly increase the absolute magnitude of SNe Ia by $\Delta$$M_{B}$ $\backsimeq$ 0.2 \cite{2021PhRvD.104b1303M,2021PhRvD.103h3517A}.

Using the publicly available SH0ES data described in \citet{2021ApJ...908L...6R}, the extended analyses were given by \citet{2022Univ....8..502P} in a detailed and comprehensive manner. They found that when an absolute magnitude $M_{B}$ transition of the SNe Ia at $D_{c}$ $\simeq$ 50 Mpc (about 160 Myrs ago) can drop, the $H_{0}$ constraint drops from 73.04 $\pm$ 1.04 km/s/Mpc to 67.32 $\pm$ 4.64 km/s/Mpc, which is in full consistency with the Planck results. When the inverse distance ladder constraint on $M_{B}^{>}$ is included in the analyses, the uncertainties for $H_{0}$ reduce dramatically ($H_{0}$ = 68.2 ± 0.8 km/s/Mpc) and the $M_{B}^{>}$ transition model is strongly preferred over the baseline SH0ES model in terms of the Akaike Information Criterion (AIC) \cite{10.1214/aos/1176344136} and the Bayesian Information Criterion (BIC) \cite{1100705} model selection criteria. Similar hints for a transition behavior is found for the other three main parameters of the analysis ($b_W$, $M_W$ and $Z_W$) at the same critical distance $D_{c}$ $\simeq$ 50 Mpc, even though in that case the $H_{0}$ estimation is not significantly affected \cite{2022Univ....8..502P}. In addition, \citet{2022MNRAS.515.2790W} also reanalysed the SNe Ia and Cepheids' observations and found that the $H_{0}$ local measurements become dependent on the choice of SN reference colour. These recent investigations hint towards the need of more detailed Cepheid + SNe Ia calibrating data at distances $D_{c}$ $\gtrsim$ 50Mpc, i.e., at the high end of rung 2 on the distance ladder.

\begin{figure}[H]
\includegraphics[width=7 cm]{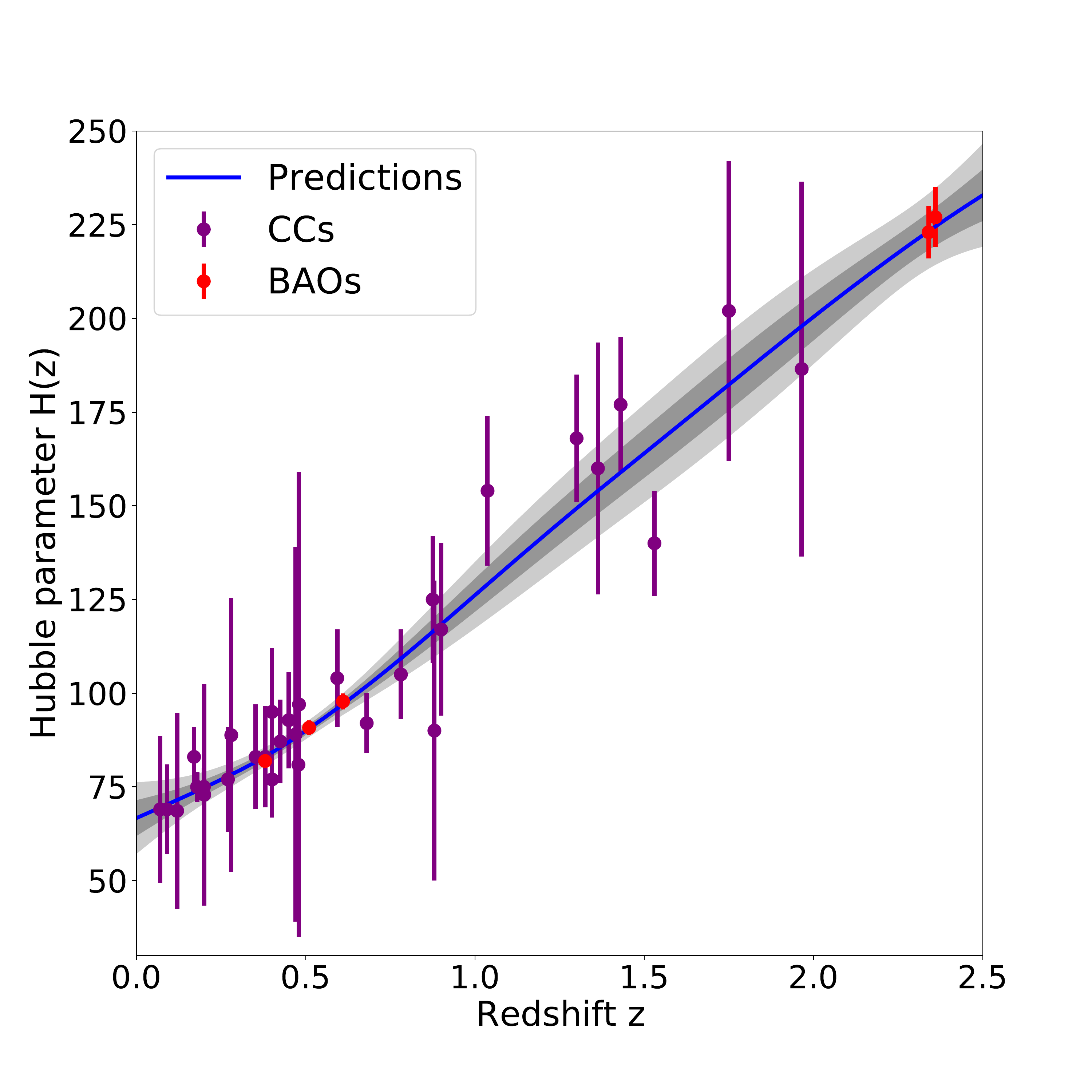}
\caption{Smoothed H(z) function (blue solid line) with 2$\sigma$ errors (gray regions) obtained from the 36 H(z) data (31 CCs + 5 BAOs) employing GP method. GP regression is implemented by employing the package \emph{scikit-learn} (https://scikit-learn.org) \cite{scikit-learn} in the Python environment. (Source: Figure~4 in \citet{2021MNRAS.507..730H}.) 
\label{fig5}}
\end{figure}

In a recent analysis, \citet{2021PhRvD.103j3509K} construct the $H_{0}$ diagnostic $\mathcal{H}0(z)$:
\begin{linenomath}
\begin{equation}\label{eq:H0D}
\mathcal{H}0(z) = \frac{H(z)}{\sqrt{1-\Omega_{m0} + \Omega_{m0}(1 + z)^{3}}},
\end{equation}
\end{linenomath} 
to specify the 
possible deviations from the flat $\Lambda$CDM model using the Gaussian process (GP) method \cite{scikit-learn}. The Gaussian process has been extensively used for cosmological applications, such as the constraint on $H_{0}$ \cite{2018JCAP...04..051G,2018ApJ...856....3Y,2019ApJ...886L..23L,2020ApJ...895L..29L,2022PhyS...97h5011S} and the comparison of cosmological models \cite{2018JCAP...02..034M}. A more detailed explanation can be discovered from the literature \cite{2006gpml.book.....R,2018arXiv180702811F,SCHULZ20181}. As shown in Figure~\ref{fig5}, we are given the reconstructed result of 36 H(z) data (31 CCs + 5 BAOs) using the GP method. From this figure, it can be learned that combining the H(z) data and GP method allows us to obtain a continuous function \emph{f(z)} to represent the discrete H(z) data. Utilizing the function \emph{f(z)}, the H(z) value at any redshift within a certain range can be obtained, including $H_{0}$ at $z$ = 0. The effective range mainly depends on the highest redshift of the H(z) data used. Finally, H(z) follows from the continuous function \emph{f(z)} and the $\Omega_{m0}$ is the 
adopted Planck values \cite{2020A&A...641A...6P}. Based on the H(z) data and GP method, they found a running $H_{0}$ with redshift $z$. The main result can be found from Figure~2 in \citet{2021PhRvD.103j3509K}.

\begin{figure}[H]
\includegraphics[width=6.8 cm]{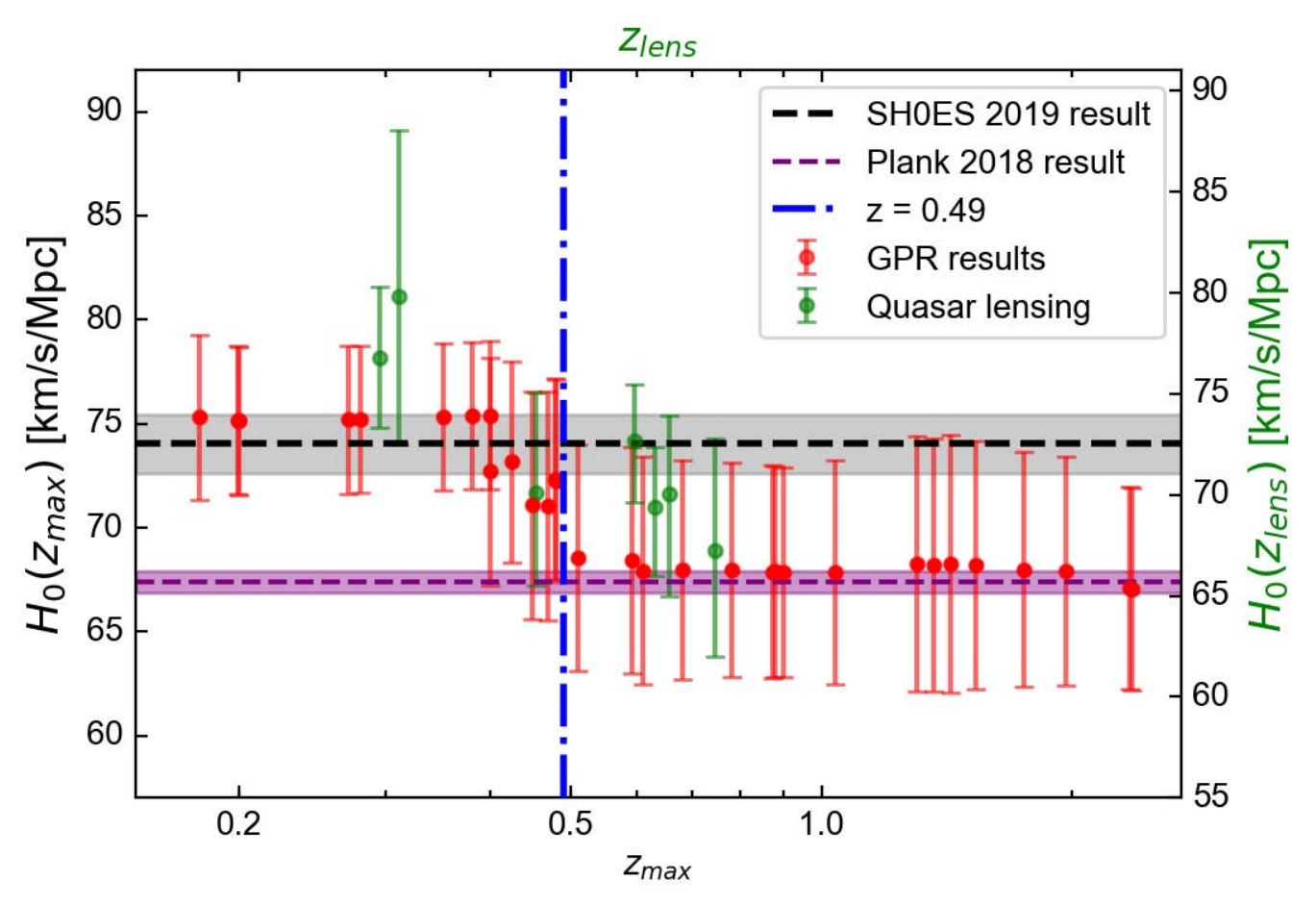}
\includegraphics[width=6.8 cm]{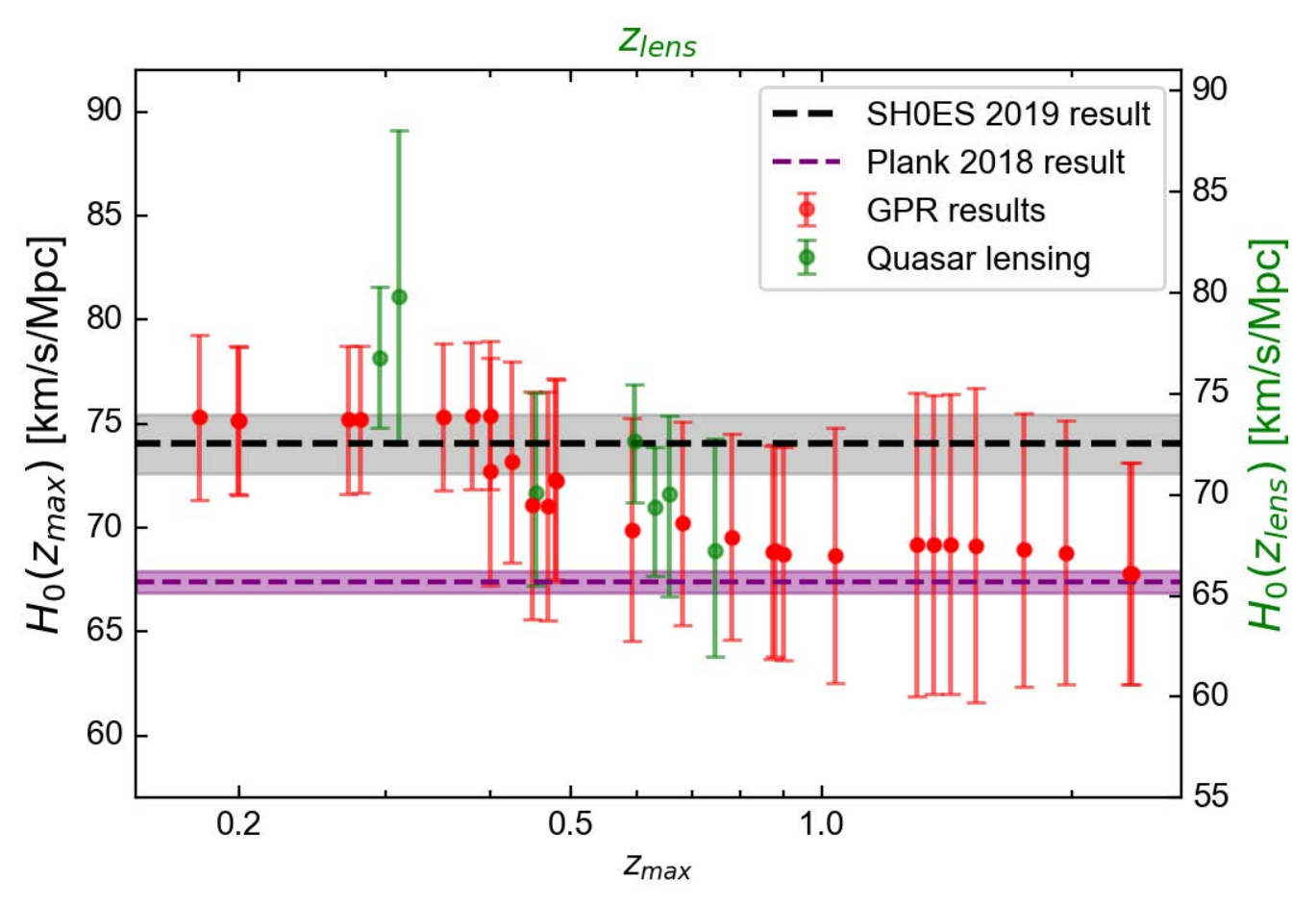}
\caption{Predictions of $H_{0}(z_{max})$ adopting the Mat\'{e}rn kernel from the 36 $H(z)$ data (31 CCs + 5 BAOs) binned by the cumulative method. $H_{0}(z_{max})$ is the $H_0$ value derived from a data-set with maximal redshift $z_{\rm max}$. Red points are the predictions of $H_{0}(z_{max})$ based on our analyses. The gray and purple regions correspond to the results of SH0ES and \emph{Planck} collaborations. Blue dotted line (\mbox{$z$ = 0.49}) is the transition redshift. We also show the $H_0$ results derived from quasar lens observations with green points in ($z_{lens}$, $H_{0}(z_{lens})$) coordinates. (Source: Figures~3 and 4 in \citet{2022MNRAS.517..576H}.)
\label{fig6}}
\end{figure}

Also employing the $H(z)$ data \cite{2018ApJ...856....3Y} and the GP method, \citet{2022MNRAS.517..576H} reported a late-time transition of $H_{0}$, i.e., $H_{0}$ changes from a low value to a high one from an early to late cosmic time that can be used to relieve the $H_{0}$ tension. Unlike previous studies \cite{2020PhRvD.102j3525K,2021ApJ...912..150D}, they processed the H(z) data using a cumulative binning method. An introduction to this method can be found in Figure~1 and Section 2 in \citet{2022MNRAS.517..576H}. They found that the redshift of the $H_{0}$ transition occurs at $z \sim 0.49$. Without proposing a new cosmological model, their finding can be used to relieve the $H_{0}$ tension with a mitigation level of around 70 percent, 
 and is consistent with the H0LiCOW results in the 1$\sigma$ range. They also tested the influence of BAOs on the result, and concluded that removing the BAOs data had no substantial effect, i.e., did not make the $H_{0}$ transition disappear. Their final results are shown in Figure~\ref{fig6}. 

\vspace{-16pt}
\vspace{-16pt}
\vspace{0.8em}
\unskip
\vspace{0.8em}

Recently, utilizing the latest SNe Ia sample (Pantheon+ sample) and H(z) data, \citet{2022arXiv221200238J} presented a novel non-parametric method to estimate $H_{0}$ as a function of the redshift. They found a descending trend of $H_{0,z}$ with the statistical significance of 3.6$\sigma$ and 5.1$\sigma$, corresponding to the equal-number and equal-width binning methods, respectively. Here, $H_{0,z}$ defined as the value of $H_{0}$ are derived from the cosmic observations at redshift $z$. The main results are presented in Figures~1 and 2 of \citet{2022arXiv221200238J}. The evolution of $H_{0,z}$ can effectively relieve the Hubble tension without the early-time modifications. Moreover, the results of the AIC and BIC demonstrate that the observational data favor the $H_{0,z}$ model over the \mbox{$\Lambda$CDM model}. Recently, utilizing a different approach than \citet{2022arXiv221200238J}, \citet{2023arXiv230112725M} also found a similar $H_{0}$ descending trend from the Pantheon+ sample.

The statistical signification of the $H_{0}$ descending trend found from the quasar lensing is not high, at only 1.7$\sigma$. Moreover, it is not clear that the $H_{0}$ descending trend is not caused by systematics. This still provides a new diagnostic for the $H_{0}$ tension. The descending trend of $H_{0}$ has also been discovered by utilizing the different data-sets and methods, most of which are based on an explicit model ($\Lambda$CDM or $w_{0}w_{a}$CDM model) \cite{2020PhRvD.102j3525K,2020MNRAS.498.1420W,2021ApJ...912..150D,2021PhRvD.103j3509K,2022arXiv220611447C,2022Galax..10...24D,2022A&A...668A..34H,2022arXiv221200238J,2022PhRvD.106d1301O}, except for \citet{2022MNRAS.517..576H}. The $H_{0}$ descending trend can effectively alleviate the $H_{0}$ tension. If this trend is substantiated going forward, a late-time modification consistent with most observations is required. However, some studies are not in favor of modifying the late-time universe \cite{2021CQGra..38r4001K,2022arXiv220608440K}. Among many late-time solutions, local void \cite{2013ApJ...775...62K,Wang2013,2022CQGra..39r4001C,2022MNRAS.511L..82Y}, modified gravity \cite{2020FoPh...50..893C,2021MNRAS.500.1795B} and modified cosmological models \cite{2019PDU....2600385D} might be considered as competitive candidates. The local void model has been disfavored by the SNe Ia data \cite{2019ApJ...875..145K,2020MNRAS.491.2075L,2021PhRvD.103l3539C}, but can not completely be ruled out. Of course, there is also some evidence supporting the existence of the local void model \cite{2020A&A...633A..19B,2022MNRAS.515.4417K,2022arXiv220113384K}. The reasons for the transition of the Cepheid parameters and the $M_{B}$ transition of SNe Ia are still unclear.  Such a transition may be attributed to either new physics or to unknown systematics hidden in the data \cite{2021PhRvD.104l3511P}. If the source of the demonstrated transitions are physical, it could lead to new cosmological physics beyond the $\Lambda$CDM model. In previous studies, there are precedents, e.g., the gravitational constant in the context of a recent first-order phase transition \cite{1977PhRvD..15.2929C,1977PhRvD..16.1762C,2014PhRvD..90f3009P} to a new vacuum of a scalar-tensor theory, or in the context of a generalization of the symmetron screening mechanism \cite{2022PhRvD.106d3528P}. A similar first-order transition was implemented in early dark energy models \cite{2020PhRvD.102f3527N}, attempting to change the last scattering sound horizon scale without affecting other well-constrained cosmological observables. Thus, even though no relevant detailed analysis has been performed so far, there are physical mechanisms that could potentially induce the SNe Ia luminosity transition degree of freedom. In any case, in the face of these evidences that may be new physics beyond the $\Lambda$CDM model, one should not just be skeptical and do nothing.

\section{Conclusions and Future Prospects}\label{sec6}
The $\Lambda$CDM model as the current standard cosmological model is consistent with almost all of the observational probes available until the present. However, it is not perfect, and there are still many theoretical difficulties and tensions. The significant discrepancy between the $H_0$ values measured from the local distance ladder and from the cosmic microwave background, i.e., Hubble tension, is the most serious challenge to the standard $\Lambda$CDM model. In this review, we have revisited this as the hottest issue, incorporating the \mbox{latest research}.

Until now, there has been a 4-6$\sigma$ discrepancy in $H_{0}$ measured by these two approaches, and the discrepancy is still increasing (see Sections \ref{sec1} and \ref{sec2} and the reference therein). Initially, possible systematics in the Planck observations and the HST measurements were thought to be responsible for the $H_{0}$ tension. However, this possibility has been largely ruled out \cite{2019ApJ...876...85R,2017A&A...607A..95P,2018ApJ...867..108J,2019MNRAS.484L..64S,2020A&A...641A...1P,2020A&A...641A...7P,2020A&A...644A.176R,2022ApJ...934L...7R,2022MNRAS.514.4620D}. The current arbitration results given by other independent observations (including quasar lensing, Megamaser, GW, FRB, TRGB, etc.) cannot effectively arbitrate the $H_{0}$ tension. See Section \ref{sec3} for details. Many researchers therefore choose to believe that the Hubble tension may be caused by new physics beyond the $\Lambda$CDM model \cite{2020PhRvD.102b3518V}. So far, there have been a lot of schemes proposed to solve the $H_{0}$ tension (see \citet{2021CQGra..38o3001D,2021A&ARv..29....9S,2022NewAR..9501659P} for a review). We have reviewed the classifications of solving schemes for the $H_{0}$ tension based on previous work by \citet{2021CQGra..38o3001D} and \citet{2021A&ARv..29....9S} as well as newer studies (see Section \ref{sec41}). According to the research and discussions about solving the $H_{0}$ tension, we tend to divide all of the solutions into two categories: sequential and reverse-order schemes. In Section \ref{sec5}, we mainly review the reverse-order schemes, which might hint at new physics beyond the $\Lambda$CDM. Some of these schemes have discovered the late-time $H_{0}$ descending trend \cite{2020PhRvD.102j3525K,2020MNRAS.498.1420W,2021ApJ...912..150D,2021PhRvD.103j3509K,2022arXiv220611447C,2022Galax..10...24D,2022A&A...668A..34H,2022arXiv221200238J,2022PhRvD.106d1301O} or the late-time $H_{0}$ transition \cite{2022MNRAS.517..576H} that can be used to alleviate the $H_{0}$ tension through different independent observations and different methods. The remaining schemes found that considering the new degrees of freedom (parameter transition) can also effectively alleviate the $H_{0}$ tension when analyzing the Cepheid and SNe Ia data \cite{2021PhRvD.104l3511P,2022Univ....8..502P,2022MNRAS.515.2790W}.

Looking to the future, two new CMB Stage 4 telescopes will be operational in Chile and the South Pole around 2030, which will further extend the spectral resolution. The depth of these surveys will be able to support or rule out many precombination modifications based on the $\Lambda$CDM model. The Zwicky Transient Facility and Foundation surveys of nearby SNe Ia will effectively reduce the potential calibration issues of the local distance ladder by resolving the underlying population characteristics, having cleaner selection functions, and providing more galaxies. The $H_{0}$ arbitration of independent observations will also be improved. The James Webb Space Telescope (JWST) will greatly expand the range of TRGB observation, and provide continuity in the case of any further degradation of the ageing HST. The VIRGO detector in Italy, and recently the KAGRA detector in Japan will provide more frequent event detections and better sky localization. They 
 will provide a 2$\%$ measurement of $H_{0}$ within this decade by combining with an improved instrumental calibration. Many thousands of lenses will be detected by the wide-field surveys, such as the Vera Rubin Observatory, Euclid and the Nancy Grace Roman Observatory, and hundreds of which will have accurate time delay measurements \cite{2010MNRAS.405.2579O,2015ApJ...811...20C,2019A&A...631A.161H}. Thus, there is a strong incentive to resolve the remaining systematics in the modelling and speed up the analysis pipeline, then clear the relationship between the $H_{0}$ descending trend and systematics. The ASKAP and Very Large Array will provide a large number of positioned FRBs in the future, which will provide higher-precision $H_0$ measurements. In addition, the e-ROSITA all-sky survey, French--Chinese satellite space-based multi-band astronomical variable objects monitor (SVOM) \cite{Wei2016}, Einstein Probe (EP) \cite{Yuan2015}, and Transient High-Energy Sky and Early Universe Surveyor (THESEUS) \cite{Amati2018} space missions together with ground- and space-based multi-messenger facilities will allow us to investigate $H_{0}$ tension in the poorly explored high-redshift universe.

\vspace{6pt}
\authorcontributions{Conceptualization, J.P. and F.Y.; investigation, J.P.; writing---original draft preparation, J.P.; writing---review and editing, J.P. and F.Y.; supervision, F.Y.; All authors have read and agreed to the published version of the manuscript.}

\funding{This work was supported by the National Natural Science Foundation of China (grant No. U1831207), the China Manned Spaced Project (CMS-CSST-2021-A12), National Natural Science Foundation of China (grant No. 12273009), Jiangsu Funding Program for Excellent Postdoctoral Talent (20220ZB59) and Project funded by China Postdoctoral Science Foundation (2022M721561).}

\dataavailability{Not applicable.} 

\acknowledgments{We thank the anonymous referee for constructive comments. This work was supported by the National Natural Science Foundation of China (grant No. U1831207), the China Manned Spaced Project (CMS-CSST-2021-A12), National Natural Science Foundation of China (grant No. 12273009), Jiangsu Funding Program for Excellent Postdoctoral Talent (20220ZB59) and Project funded by China Postdoctoral Science Foundation (2022M721561).}

\conflictsofinterest{The authors declare no conflict of interest.} 


\begin{adjustwidth}{-\extralength}{0cm}

\reftitle{References}

\PublishersNote{}
\end{adjustwidth}
\end{document}